\documentclass[11pt,a4paper]{article}
\pdfoutput=1
\usepackage{jheppub}
\usepackage{graphicx} 
\usepackage{epstopdf} 
\usepackage{dcolumn} 
\usepackage{xcolor} 
\usepackage{amsmath,amssymb} 
\usepackage{hyperref} 
\usepackage[utf8]{inputenc} 
\usepackage[english]{babel} 
\usepackage[displaymath, mathlines]{lineno} 
\usepackage{blindtext} 
\usepackage{ifthen} 
\usepackage{orcidlink} 
\usepackage{placeins}
\usepackage{multirow}
\usepackage{arydshln} 

\graphicspath{{figures/}} 

\newboolean{articletitles}
\setboolean{articletitles}{true} 

\input{belle2-symbols}

\newcommand{\Btagm}{B_{\mathrm{tag}}}

\begin{document}

\vspace*{-3\baselineskip}
\resizebox{!}{2cm}{\includegraphics{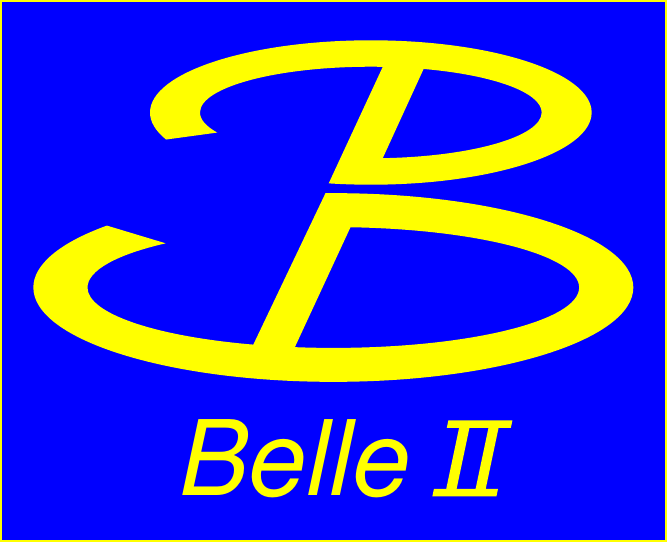}}
\begin{flushright}
Belle II preprint: 2025-014\\
KEK preprint: 2025-13\\
\end{flushright}

\title{Search for lepton flavor-violating decay modes \boldmath{$B^0 \to K^{\ast 0}\tau^\pm\ell^\mp$  ($\ell = e,\mu$)} with hadronic B-tagging at Belle and Belle II
}
\collaboration{The Belle and Belle II Collaborations}
  \author{I.~Adachi\,\orcidlink{0000-0003-2287-0173},} 
  \author{Y.~Ahn\,\orcidlink{0000-0001-6820-0576},} 
  \author{H.~Aihara\,\orcidlink{0000-0002-1907-5964},} 
  \author{N.~Akopov\,\orcidlink{0000-0002-4425-2096},} 
  \author{S.~Alghamdi\,\orcidlink{0000-0001-7609-112X},} 
  \author{M.~Alhakami\,\orcidlink{0000-0002-2234-8628},} 
  \author{A.~Aloisio\,\orcidlink{0000-0002-3883-6693},} 
  \author{K.~Amos\,\orcidlink{0000-0003-1757-5620},} 
  \author{M.~Angelsmark\,\orcidlink{0000-0003-4745-1020},} 
  \author{N.~Anh~Ky\,\orcidlink{0000-0003-0471-197X},} 
% ...

\abstract{We present the results of a search for the charged-lepton-flavor violating decays $B^0 \rightarrow K^{*0}\tau^\pm \ell^{\mp}$, where $\ell^{\mp}$ is either an electron or a muon.
The results are based on 365 \invfb and 711 \invfb datasets collected with the Belle II and Belle detectors, respectively.
We use an exclusive hadronic $B$-tagging technique, and search for a signal decay in the system recoiling against a fully reconstructed $B$ meson. 
We find no evidence for $B^0 \rightarrow K^{*0}\tau^\pm \ell^{\mp}$ decays and set  upper limits on the branching fractions in the range of (2.9--6.4)$\times10^{-5}$ at 90\% confidence level.
}

\maketitle
\flushbottom

\section{Introduction}
\label{sec:intro}

In the standard model (SM) of particle physics charged lepton flavor violating (LFV) processes are highly suppressed. They can occur only via neutrino-mixing with rates far below the current and foreseen experimental reach~\cite{Calibbi:LFV}. 
However, in many extensions beyond the standard model (BSM), LFV decays are enhanced since they are not protected by any fundamental symmetry (see Ref.~\cite{Davidson:2022jai} for a recent review). 
An observation of such decays would thus provide indisputable evidence of physics beyond the SM.
In addition, in the SM, lepton flavor universality (LFU) holds, meaning that the interaction of the three generations of leptons with the gauge bosons is identical, except for differences arising from their masses~\cite{Weinberg:EWK,GIM:EWK}.
The  recent anomalies observed in $b\to c \tau \nu$ transitions~\cite{LHCb:RD, HFLAV:23} may nevertheless suggest deviations from LFU, which also imply LFV in many BSM scenarios. 
For instance, a heavy $Z'$ boson mediator or a leptoquark-mediated transition would produce LFV~\cite{theory:LFVLQ2016, theory:LFVLQ2021, theory:LFVZprime}. 
Measurements of $b\to s$ transitions can also be used to investigate LFU deviation and LFV~\cite{Becirevic:2016zri, theory:LFVbsll24, LHCB:sll1, LHCB:sll2}.
In particular, the recent excess observed by Belle~II in $b\to s\nu\bar \nu$ transitions~\cite{Belle2:Knunu} can be described by allowing LFV~\cite{Theory:Knunu_LFV}, which could give an enhancement of the branching fractions of $B\to K\tau^\pm\ell^\pm$ up to $3\times 10^{-6}$, where $\ell$ stands for $e$ or $\mu$. 
This is close to the current experimental limits and motivates further searches. 

The LHCb experiment searched for $B^0\rightarrow K^{*0}\tau^- \mu^+$ ($B^0\rightarrow K^{*0}\tau^+ \mu^-$) decays using a 9~\invfb dataset~\cite{LHCb:2022wrs} and obtained upper limits of  0.82 (1.0)$\times 10^{-5}$ at the 90\% confidence level (C.L.).
Searches for $B^+ \rightarrow K^+\tau^{\pm}\ell^{\mp}$ have been also performed by the BaBar \cite{BaBar:2012} and Belle \cite{Belle:2023} and LHCb \cite{LHCb:B2Kmutau2020} experiments, setting the best upper limits in these modes between 0.6 and 2.5$\times10^{-5}$ at 90\% C.L, as well as searches for $B^0_s \rightarrow \phi \mu^\pm \tau^\mp$ by LHCb \cite{LHCb:Bs2phimutau2024} with upper limits at $10^{-5}$ level.
Here, we present the first search for $B^0 \rightarrow K^{*0}\tau^\pm e^\mp$ decays\footnote{An analysis has been performed in parallel by the LHCb experiment on $B^0 \rightarrow K^{*0}\tau^+ e^-$ and $B^0 \rightarrow K^{*0}\tau^- e^+$ LFV decays, with preliminary upper limits set at $4.9\times10^{-6}$ and $5.9\times10^{-6}$, respectively \cite{LHCb:B2Ksttaue2025}.} and the first search for $B^0 \rightarrow K^{*0}\tau^\pm \mu^\mp$ decays at a $B$ factory, using the combined dataset of the Belle and Belle~II experiments, with integrated luminosities of 711~\invfb and  365~\invfb, respectively adding up to a total of 1076~\invfb. 
Four different final states are distinguished according to the flavor of the final state lepton  $\ell$, and to the sign of its charge with respect to that of the kaon from the $K^{*0}$: 
same-sign $SS\ell$ for $B^0 \rightarrow K^{*0}(\to K^+\pi^-)\tau^- \ell^+$
and opposite-sign $OS\ell$ for $B^0 \rightarrow K^{*0}(\to K^+\pi^-)\tau^+ \ell^-$.
Charge conjugated final states are implied through this paper. 

We use a $B$-tagging technique to reconstruct one $B$ meson decaying into hadronic modes.
The signal is then searched for in the system recoiling against the $K^{*0}\ell$ from the other $B$ meson of the pair and the fully reconstructed tagged $B$, and representing the $\tau$ lepton candidate.
Combinatorial and some targeted backgrounds are rejected using a cut-based approach followed by a Boosted Decision Tree (BDT).
Finally the signal is extracted from a simultaneous fit to the $\tau$ lepton recoil mass in the Belle and Belle II datasets.
\section{The Belle and Belle II detectors, simulation and data samples}
\label{sec:exp}
The Belle~II experiment is located at SuperKEKB~\cite{Akai:2018mbz, SKEKB:accDesign}, an accelerator colliding electrons and positrons with center-of-mass energies near the $\Upsilon(4S)$ resonance. 
The $\belletwo{}$ detector~\cite{Abe:2010gxa} has a cylindrical geometry surrounding the interaction point and includes a two-layer silicon-pixel detector~(PXD) surrounded by a four-layer double-sided silicon-strip detector~(SVD)~\cite{Belle-IISVD:2022upf} and a 56-layer central drift chamber~(CDC). 
These detectors reconstruct the trajectories (tracks) of charged particles and measure energy loss due to ionization in the material of the detector.
Only one sixth of the second layer of the PXD had been installed for the data analyzed here.   
Surrounding the CDC are a time-of-propagation detecter~(TOP)~\cite{Kotchetkov:2018qzw} in the central region and an aerogel-based ring-imaging Cherenkov detector~(ARICH) in the forward region.  
These detectors provide information used to identify charged particles.  
Surrounding the TOP and ARICH is an electromagnetic calorimeter~(ECL) based on CsI(Tl) crystals providing energy and timing measurements, primarily for photons and electrons.
Outside the ECL is a superconducting solenoid magnet that provides a 1.5~T axial magnetic field. 
The magnetic flux is returned via an iron yoke, which serves the dual purpose of also being instrumented with resistive-plate chambers and plastic scintillator modules (KLM) to detect muons, $K^0_L$ mesons, and neutrons. 
The symmetry axis of the magnet, which almost coincides with the direction of the electron beam, defines the $z$ axis.

The Belle detector was located at the interaction point of the KEKB collider~\cite{Kurokawa:2001nw, Abe:2013kxa}.  
It shares a similar structure to $\belletwo{}$, but lacks a silicon pixel detector and plastic scintillators in the KLM, uses aerogel threshold Cherenkov counters (ACC) and a barrel-like arrangement of time-of-flight scintillation counters (TOF) for particle identification.
Vertexing and tracking are performed using the Belle SVD and CDC. A detailed description of the Belle detector can be found in Ref.~\cite{ABASHIAN2002117, Belle:2012iwr}.

This analysis uses the 711\invfb Belle dataset corresponding to $(771.6\pm10.6)\times10^6$ $\Upsilon(4S)$ events, and the dataset collected by Belle II during the first data taking period, corresponding to 365\invfb or $(387.1\pm5.6)\times10^6$ $\Upsilon(4S)$ events.

Monte Carlo (MC) simulated events are used to optimize the signal selection, to improve background rejection, to model the signal and measure its efficiency as well as to estimate the systematic uncertainties.
The signal \BKstTauell channels are modeled using a uniform three-body phase space model; 20 million events  are produced for each decay mode. 
Simulated samples that reproduce the composition of Belle and Belle II events, including $B\bar{B}$ and $e^+e^- \rightarrow \qqbar$ continuum (where $q$ indicates an $u$, $d,$ $s$ or $c$ quark) backgrounds, and equivalent to approximately four times the data luminosity are used to investigate the sample composition and validate the analysis before examining the signal region in data.
Simulated events are generated using the \texttt{KKMC} generator for quark-antiquark production from $e^+e^-$ collisions~\cite{Jadach_2000}, the \texttt{PYTHIA8} (\texttt{PYTHIA6} for Belle) generator for hadronization~\cite{Sjostrand:2014zea, Sj_strand_2006}, the \texttt{EvtGen} software package and the \texttt{PYTHIA} generator for the decay of the generated hadrons~\cite{Lange:2001uf}, the \texttt{PHOTOS} package for the final state radiation (FSR)~\cite{Barberio:1990ms} and the \texttt{Geant4} (\texttt{Geant3} for Belle) software package for the detector response~\cite{Agostinelli:2002hh, Brun:1994aa}. 
The simulation includes beam-induced background data overlay~\cite{Liptak2022}.
The data and the MC simulations are processed using the Belle II analysis software (basf2)~\cite{Kuhr:2018lps, basf2-zenodo} and Belle data and MC are converted from the Belle analysis software (basf)~\cite{Belle:basf} format into the Belle II format for basf2 compatibility using the \texttt{B2BII} framework~\cite{Gelb:2018agf}.

\section{Event selection and background rejection}
\label{sec:selection}

Events are selected by a hardware  trigger targeting $\Upsilon(4S)\to B^0\bar B^0$ events, based on the charged-particle multiplicity and total energy, in order to suppress low-multiplicity events and beam-related background. 
We reconstruct the selected events using a $B$-tagging approach. 
We reconstruct one of the $B$ meson in the pair, called $\Btag$, in exclusive hadronic decays, using the hadronic Full Event Interpretation (FEI) $B$-tagging algorithm~\cite{Keck:2018lcd}. 
FEI is a machine-learning based algorithm developed for $B$-tagged analyses at \belle and \belletwo. 
It reconstructs $B$ meson candidates from exclusive decays, using a hierarchical approach starting from reconstructed charged and neutral particles in the detector. 
We then reconstruct the second $B$ meson of the event, called $\Bsig$, in our signal channel $B^0\rightarrow K^{*0} \tau^{\pm}\ell^{\mp}$ from the tracks left after \Btag reconstruction. 
Since the \Bsig final state contains at least one neutrino coming from the $\tau$ decay, its kinematics cannot be fully determined. 
However, the \Bsig can be constrained by exploiting the knowledge of the $\Upsilon(4S)$ initial state and information on the fully-reconstructed $\Btag$. 

The \Btag candidates are selected by requiring at least three tracks, three ECL energy deposits (clusters) in the event, and a visible energy in the center-of-mass frame of at least 4~\gev. 
These tracks are required to have a transverse impact parameter $d_0<0.5$~cm, a longitudinal impact parameter $|z_0|<2$~cm and a transverse momentum $p_T>0.1$~\gev/$c$. 
The clusters are required to be in the angular acceptance of the CDC (polar angle from 17$^\circ$ to 150$^\circ$) with energies larger than 0.1~\gev. 
The \Btag must have a beam-energy-constrained mass $M_\mathrm{bc}=\sqrt{(E_\mathrm{beam}/c^2)^2-(p_{\Btagm}/c)^2}$ larger than $5.27$ \gevcc and an energy difference $\Delta E=E_{\Btagm}-E_\mathrm{beam}$ in the range $-0.15<\Delta E<0.1$\gev.
Here, $E_\mathrm{beam}$, $E_{\Btagm}$ and $p_{\Btagm}$ are the beam energy, and the energy and momentum of the \Btag candidate in the \epem center of mass frame. 
Each reconstructed \Btag is assigned a multivariate classifier output, $\mathcal{P}_\mathrm{FEI}$, ranging from zero to one and corresponding to candidates identified as background-like and signal-like respectively. 
We require the \Btag candidate to have $\mathcal{P}_\mathrm{FEI}>0.001$. 
If multiple \Btag candidates are reconstructed in an event, the one with the highest \PFEI is retained.
After this selection, the \Btag purity is approximately 45\% (40\%) for Belle II (Belle).

The \Bsig candidates are selected by combining three tracks, corresponding to a reconstructed $K^{*0}(K^+\pi^-)$ with a charged lepton ($e,\mu$).  
The $\tau$ lepton is not explicitly reconstructed, but up to three additional tracks are permitted to be present in the event.
The kaon, pion and lepton candidates are identified as tracks not used in the \Btag reconstruction, with $d_0<0.5$~cm, $|z_0|<5$~cm, in the CDC angular acceptance, and for hadrons only, to have at least 20 hits in the CDC (in the following we refer to tracks satisfying these requirements as ``good tracks''). 
For both Belle and Belle II, kaons and pions are identified requiring an identification likelihood ratio $\mathcal{P}_{K}$, $\mathcal{P}_{\pi}> 0.6$ 
while electrons and muons candidates must satisfy $\mathcal{P}_{e}$, $\mathcal{P}_{\mu}> 0.9$.
The hadron identification likelihood uses information from the ACC, CDC, and TOF for Belle. 
For Belle II, information from all subdetectors except the PXD and SVD is used, resulting in a kaon identification efficiency of 85\% (88\% for Belle) for a pion fake rate of 5\% (8\% for Belle) and a pion identification efficiency of 90\% (91\% for Belle) for a kaon fake rate of 7\% (6\% for Belle) at the particle identification working points $\mathcal{P}_K, \mathcal{P}_{\pi}>0.6$. 
For Belle, electrons are identified using the information from the ECL, CDC and ACC, and information from the KLM only is used for muon identification.
Using the likelihood ratio requirement  $\mathcal{P}_{e} (\mathcal{P}_{\mu})> 0.9$, the lepton identification has an efficiency of 92\% (89\%) and a pion fake rate of 0.3\% (1.4\%) for electrons (muons with momentum larger than 0.6 GeV/$c$).
For Belle II, the electron identification uses a boosted-decision-tree (BDT) classifier trained with information from all sub-detectors except the PXD and SVD and the muon identification uses information from all sub-detectors except the PXD and SVD.
At the particle identification working point $\mathcal{P}_{e} (\mathcal{P}_{\mu})>0.9$, the electron (muon) identification has an efficiency of 86\% (89\%) and a pion mis-identification rate of 0.4\% (7\%).

To recover electron candidates with bremsstrahlung, we accept photons having minimum energies of 50 MeV within a 50~mrad angle of an electron track. We reconstruct $K^{*0}\to K^+\pi^-$ candidates combining a kaon and a pion of opposite charge.
The kaon, pion and lepton are fitted to a common vertex, and their kinematic information is updated according to the fit result.
The $K^{*0}$ candidate should have an invariant mass in the range $0.842 < M(K^+\pi^-) < 0.942~$GeV/$c^2$.
The presence of a $\tau$ lepton is inferred from the presence of a single good track $t_\tau$ with a charge opposite to that of the primary lepton. 
This track is not used in the signal kinematic reconstruction and is only used to reduce the background contamination.  

After the event is properly reconstructed we can proceed to the signal selection.
In order to reject background, we use properties of the rest-of-event (ROE) that correspond to good tracks and photons not used in the reconstruction of the \Bsig (i.e. $K^{*0}\ell$ system) candidate, the \Btag and the $t_{\tau}$ track. Photons are reconstructed from ECL clusters within the CDC acceptance and not associated with any tracks. 
Photon candidates must satisfy additional requirements, described in Ref.~\cite{Belle-II:2024qod}, to reduce photons from beam background. 
We select events with at most two tracks in the ROE, in order to retain 3-prong decays of the $\tau$ lepton, or potential signal candidate in which the \Btag is not correctly reconstructed, leading to partner-$B$ daughters tracks falling in the ROE. 
For Belle, we also require the total charge of the ROE to be zero.
This requirement is not applied to Belle II due to its smaller data sample, to preserve enough events for a BDT training and the fit.
At this stage, multiple candidates due to the different possible combinations of tracks reconstructed as $t_\tau$ or ROE tracks are kept.
Finally, to reduce contamination from $e^+e^- \rightarrow\qqbar$ events, we require an event sphericity larger than 0.2~\cite{Bjorken:1969wi} and the absolute value of cosine of the angle between the thrust axis of $\Btag$ and that of particles not used in $\Btag$ to be less than 0.9. 

The \Bsig momentum is equal in magnitude and opposite in direction to that of \Btag, $\vec{p}_{\Btag}$, and the \Bsig energy is equal to $E_\mathrm{beam}$ in the center-of-mass frame. Therefore the $\tau$ momentum and energy are given by:
\begin{equation}
    \vec{p}_\tau=-\vec{p}_{\Btag}-\vec{p}_{K^{*0}}-\vec{p}_\ell,~~~~~~\\
    E_\tau=E_\mathrm{beam}-E_{K^{*0}}-E_\ell.
\end{equation}
The $\tau$ mass is then reconstructed as 
\begin{equation}
    M_\tau=[m_B^2+M^2(K^{*0}\ell)-2(E_\mathrm{beam}E_{K^{*0}\ell} 
    + |\vec{p}_{\Btag}| |\vec{p}_{K^{*0}\ell}| \cos\theta ) ]^{1/2}.
\end{equation}
Here, $m_B$ is the known $B^0$ mass~\cite{ParticleDataGroup:2024cfk}; $M(K^{*0}\ell)$, $E_{K^{*0}\ell}$, $\vec{p}_{K^{*0}\ell}$
are the mass, energy, and momentum of the system composed
of the $K^{*0}$ and $\ell$, respectively; $\theta$ is the angle between
$\vec{p}_{\Btag}$ and $\vec{p}_{K^{*0}\ell}$.
All the above quantities are defined with respect to the center-of-mass frame.
Candidates having a reconstructed $M_\tau$ outside the range [1.0,\,2.5]\gevcc are discarded and the signal region is defined as [1.65,\,1.90]\gevcc.
To avoid biases, we do not examine the signal region until the analysis strategy is fixed.

Background can arise from $B^0\to K^{*0} J/\psi$ decays when the $J/\psi$ products are reconstructed as the $\ell t_\tau$ pair. 
They are removed requiring the  $\ell t_\tau$ invariant mass to be outside the range [3.05,\,3.15]\gevcc.
The background for $SS\ell$ modes ($B^0\rightarrow K^{*0}\tau^-\ell^+$) is mainly due to 
semileptonic $B$ decays such as $B\to D \ell\nu$, with $D\to K\pi\pi$. These events are vetoed by removing candidates with $M(K^{*0}t_\tau)$ in the range [1.83,\,1.90]\gevcc.
The main backgrounds in $OS\ell$ modes ($B^0\rightarrow K^{*0}\tau^+\ell^-$) are from $B\to D X$ decays where the $D$ meson decays semileptonically as $D\to K^{*0}\ell\nu$. 
Hadronic $B\to D X$ decays with $D\to K\pi\pi$ can be reconstructed as \OSmu signal if a pion is misreconstructed as a muon. 
A veto is thus applied for candidates with $M(K^{*0}\ell)$ in the range [1.83,\,1.90]\gevcc. 
In addition, according to the simulation, \qqbar processes amount for 7 to 24\% of the background in $SS\ell$ modes and 24 to 56\% of the background in $OS\ell$ modes after applying the above selection criteria.
For each mode, there is a slightly larger \qqbar contribution in Belle II than in Belle.

To reduce the remaining background, eight BDTs are trained separately for each signal mode, and for Belle and Belle II data using the \texttt{fastBDT} library~\cite{Keck:2018lcd}. 
The training is performed using simulated signal and \qqbar and $B\Bar{B}$ processes for the background. 
For each BDT, between twelve and fourteen input variables are chosen from a common set of fifteen variables, removing the ones that do not improve the performance.
The variables comprise quantities related to the signal $B$, with the $M(K^{*0}t_\tau)$ and $M(K^{*0}\ell)$ invariant masses, the energy of the lepton  and track from $\tau$, the $\chi^2$ probability of the $K^{*0}\ell$ vertex and its distance with respect to the interaction point in the transverse plane.
The BDTs also use event shape variables such as the sphericity and the modified Fox-Wolfram moments~\cite{Belle:2003fgr} to suppress the \qqbar background.
In addition, quantities characterizing the ROE are also considered: the number of tracks and clusters, the total cluster energy and the ROE momentum.
The requirements applied to the BDT outputs are optimized using the figure of merit defined as $\frac{\epsilon}{\alpha /2 + \sqrt{N}}$~\cite{Punzi:2003bu}, where $\epsilon$ is the signal efficiency and $N$ is the number of background events evaluated in the $M_\tau$ range [1.70,\,1.85]\gevcc, which corresponds to approximately 90\% signal coverage. We set $\alpha$  to 3, corresponding to $3\sigma$ signal significance.
If the Punzi-optimised selection results in a background that is too small to be fitted, the BDT selection is relaxed to the previous local maximum of the Punzi figure of merit.

After the BDT selection the average  multiplicity is 1.1 in both MC and data samples. In events with more than one candidate, one is selected randomly.  
\medskip

\begin{figure}[ht!]
    \centering
    \includegraphics[width=0.4\linewidth]{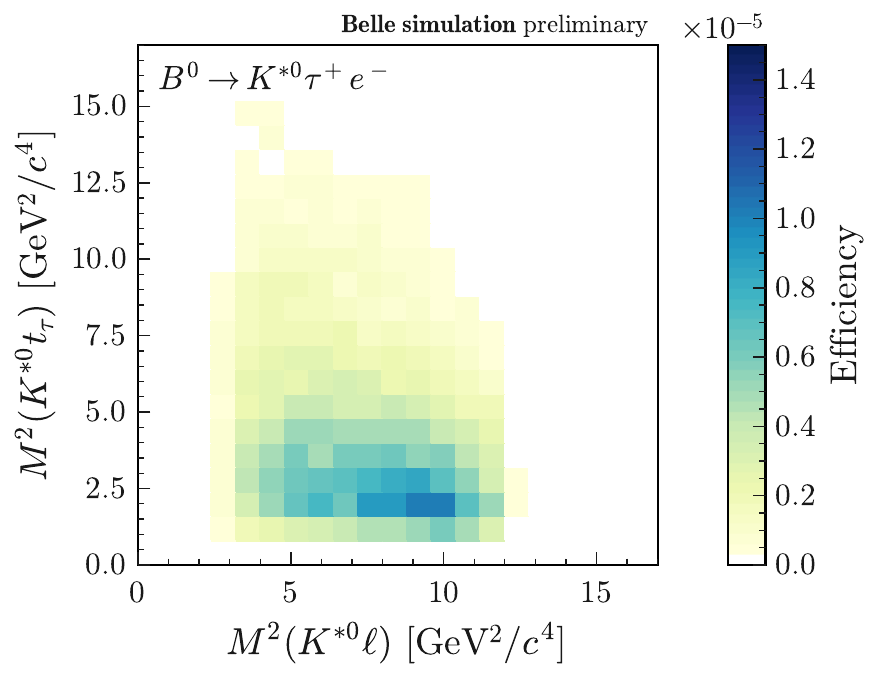}
    \includegraphics[width=0.4\linewidth]{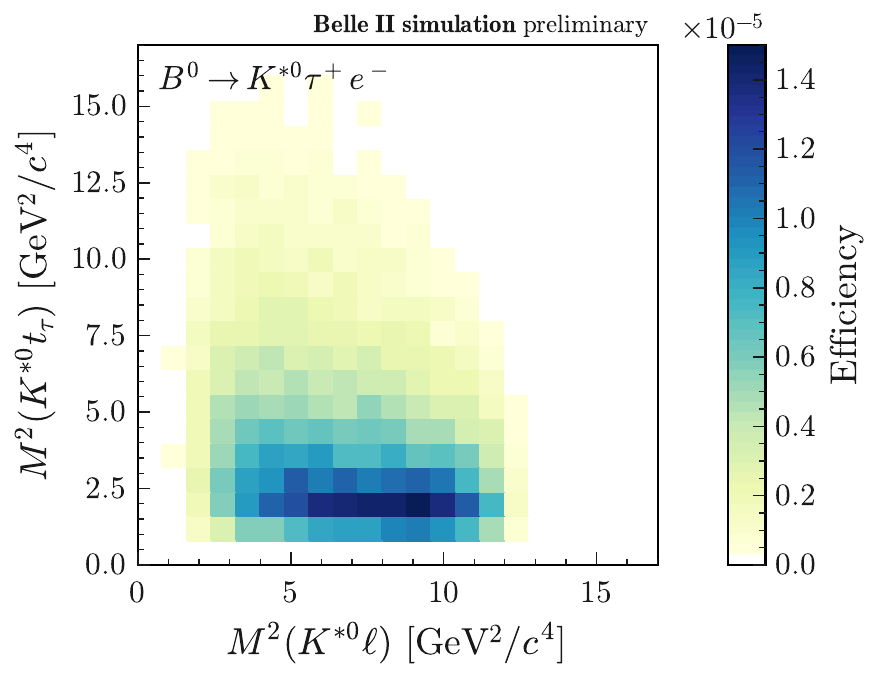}
    \includegraphics[width=0.4\linewidth]{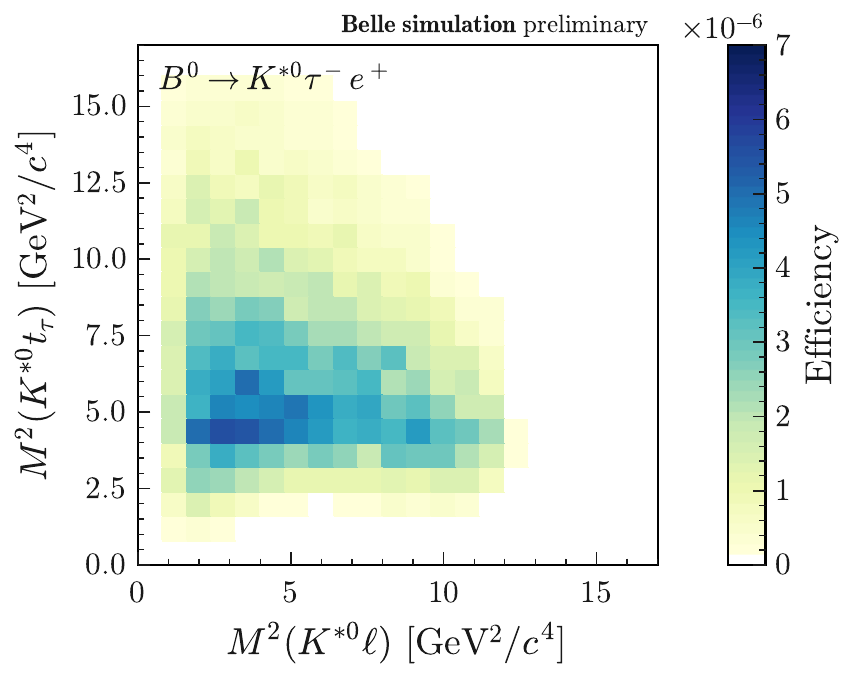}
    \includegraphics[width=0.4\linewidth]{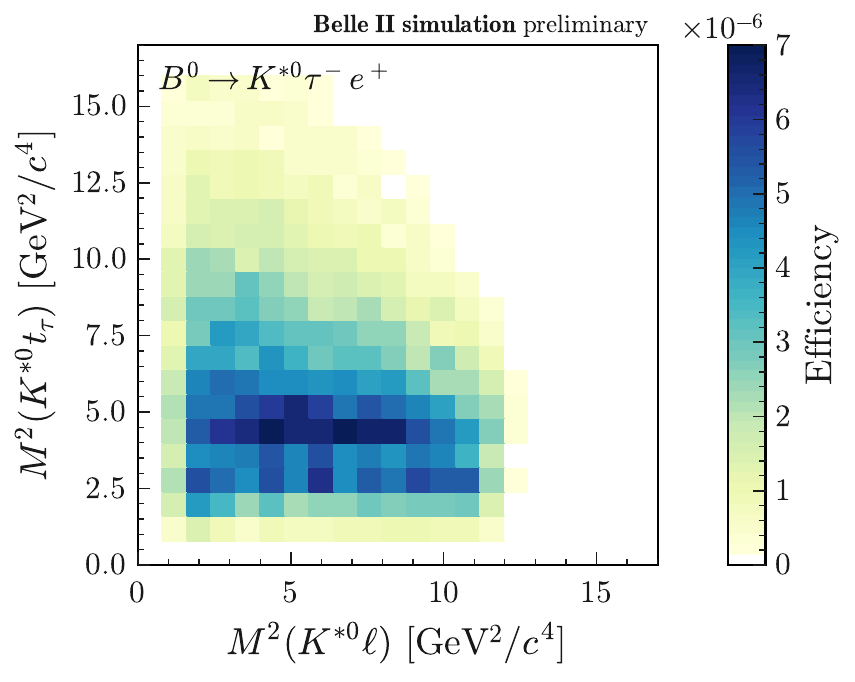}
    \includegraphics[width=0.4\linewidth]{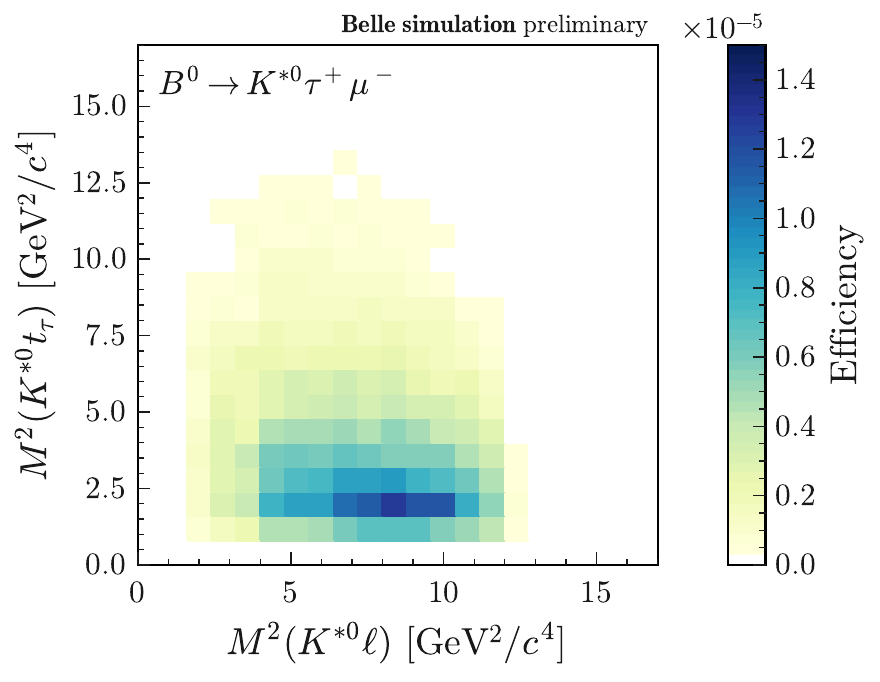}
    \includegraphics[width=0.4\linewidth]{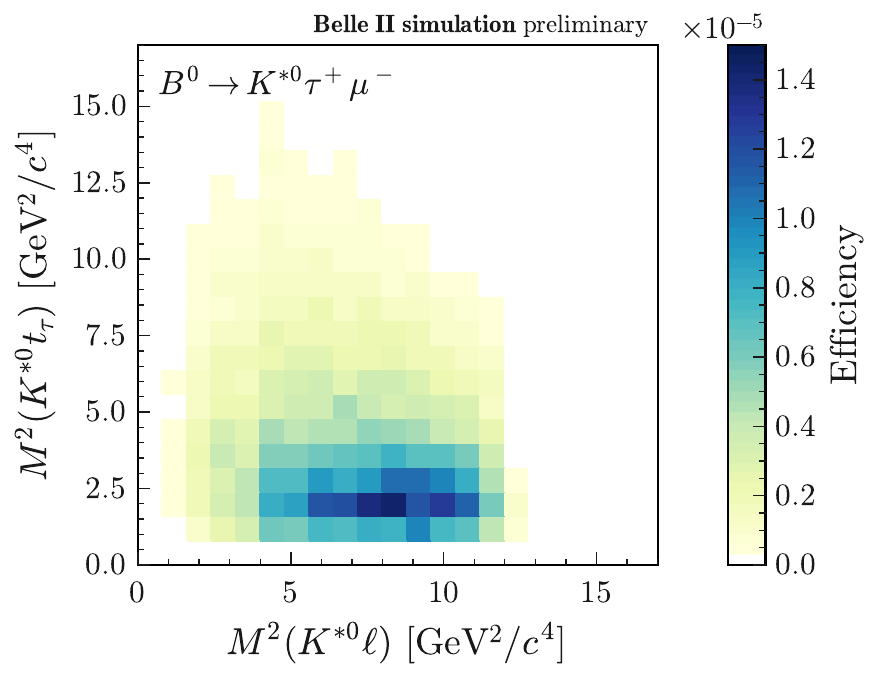}
    \includegraphics[width=0.4\linewidth]{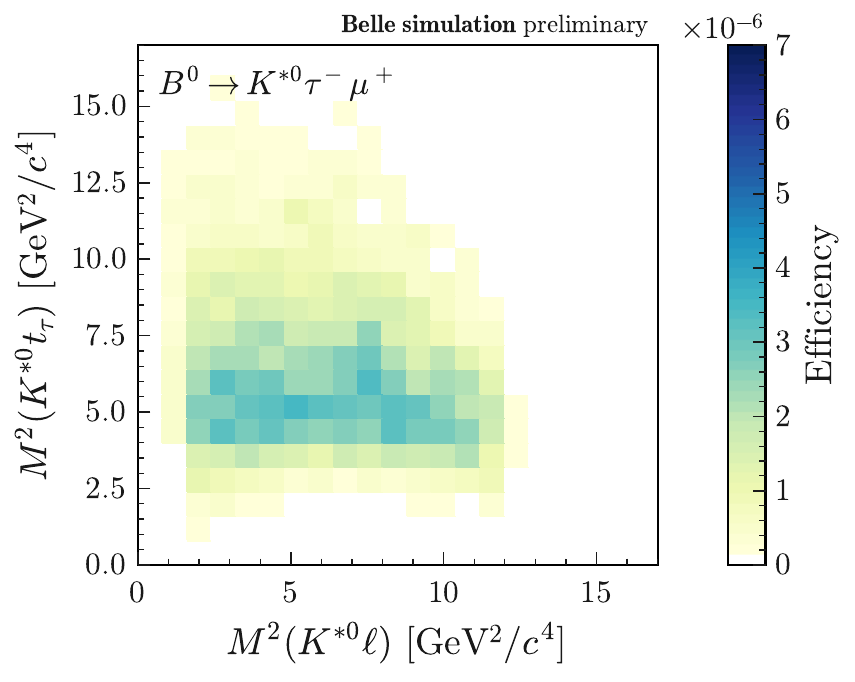}
    \includegraphics[width=0.4\linewidth]{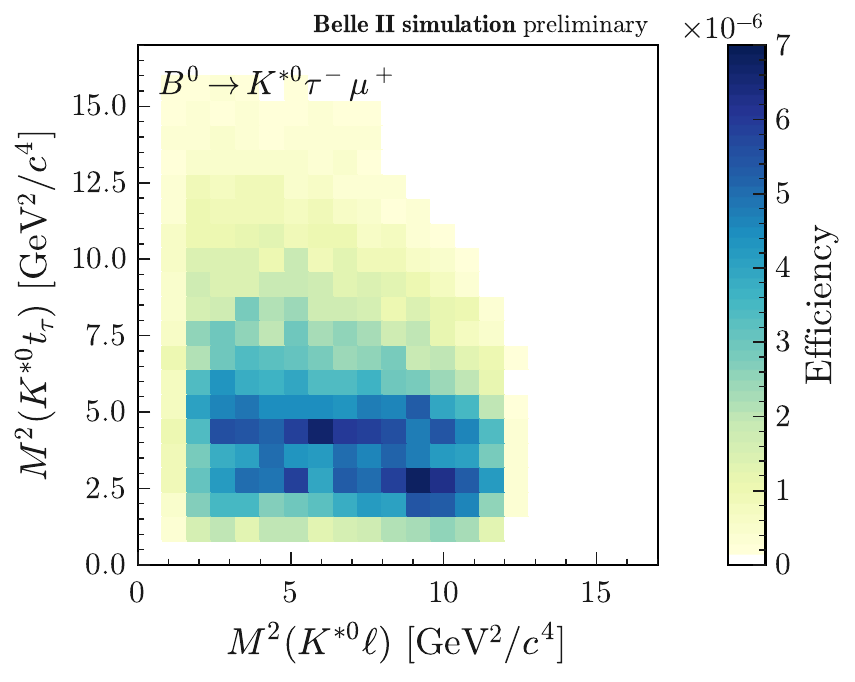}
    
    \caption{Final signal efficiencies after all selection described in sec.~\ref{sec:selection} as a function of the kinematic variables $M^2(\KstEll)$ and $M^2(\Kstarz t_{\tau})$ for Belle (left) and Belle II (right). From top to bottom: \OSe, \SSe, \OSmu, \SSmu.}
    \label{fig:eff-maps}
\end{figure}

The final reconstruction and selection signal efficiencies are given in Table~\ref{tab:eff}, while the two-dimensional final efficiencies as function of the squared invariant masses $M^2(K^{*0}t_\tau)$ and $M^2(K^{*0}\ell)$ are shown in Fig.~\ref{fig:eff-maps}. 
A uniform Dalitz distribution of the $K^{*0}\tau\ell$ system is assumed in signal generation.
The efficiencies are determined on simulated samples taking into account known mismodeling affecting the particle identification and \Btag reconstruction, detailed in Sec.~\ref{sec:syst}.
The efficiency difference between Belle and Belle II and between OS and SS modes is mainly due to the BDT selection.
On average, the optimal cut leads to a higher background rejection and thus a lower efficiency for Belle compared to Belle II, and for SS modes compared to OS modes, due to the larger \BBbar background contribution in the former.
The differences in the efficiency maps between $OS\ell$ and $SS\ell$ modes are due to the different background composition. 
In particular, $OS\ell$ modes are polluted by $B \rightarrow DX$ background where the $D$ decays semileptonically, which peaks at low values of $M(K^{*0}\ell)$. 
Those events are thus suppressed by the BDT, giving a lower efficiency in that region for $OS\ell$ final states.

\begin{table}[!hbp]
    \centering
\caption{Final signal efficiencies after all selection described in sec.~\ref{sec:selection} for each signal mode and experiment. The values contain the corrections developed to take into account known data mismodeling that affect the particle identification and the \Btag reconstruction. All values are in percent.}
\vspace{0.2cm}
\begin{tabular}{c|cccc}     
\hline
& $OSe$ & $SSe$ & $OS\mu$ & $SS\mu$\\
\hline
Belle &    0.046 & 0.038 & 0.052 & 0.024   \\
Belle II &  0.075  & 0.056 & 0.060 & 0.051 \\
\hline
\end{tabular}
\label{tab:eff}
\end{table}

\section{Branching fraction measurement}
\label{sec:fit}
The signal branching fractions are obtained from unbinned extended maximum likelihood fits to the recoil $M_\tau$ distributions in the defined fit range [1.3,\,2.3]\gevcc, simultaneously for Belle and Belle II data.
For each channel, the likelihood is expressed as
\begin{equation}
    \label{chap6 eq:pdf tot}
    \mathcal{L}(M_\tau) = \frac{e^{-(n_\text{sig}+n_\text{bg})}}{N!}\prod_{i=1}^{N}( n_\text{sig}\cdot \mathcal{P}_\text{sig}(M_\tau^i)  + n_\text{bg} \cdot \mathcal{P}_\text{bg}(M_\tau^i)) \cdot \prod_{s}\text{Gauss}(s,\sigma_s), 
\end{equation}
where $n_\text{sig}$($n_\text{bg}$) and  $\mathcal{P}_\text{sig}$($\mathcal{P}_\text{bg}$) 
are the number of events and probability density function (PDF) 
for the signal (background) and  $N$ is the total number of events. 
The Gaussian terms account for the systematic uncertainties $\sigma_s$ on the sources $s$ detailed in sec. \ref{sec:syst}.
One likelihood is defined per data sample Belle and Belle II.
We express the number of signal events in the dataset exp = Belle, Belle II as:
\begin{equation}\label{eq:BF}
    n_\text{sig}^\text{exp} = \BR(B^0\to K^{*0}\tau\ell) \times 2\times\epsilon^\text{exp} \times f^{00} \times N_{\Upsilon(4S)}^\text{exp}\times \BR(K^{*0}\to K^+\pi^-),
\end{equation} 
where $\epsilon^\text{exp}$ is the signal efficiency given in Table~\ref{tab:eff}, $f^{00}=0.4861 \pm 0.0080$ the  fraction of $\Upsilon(4S)$ decaying into $B^0\bar{B^0}$ pairs~\cite{HFLAV:23} and  $N_{\Upsilon(4S)}^\text{exp}$ is the number of produced $\Upsilon(4S)$ mesons,  $N_{\Upsilon(4S)}^\text{Belle} = (771.6\pm 10.6) \times 10^6$ and $N_{\Upsilon(4S)}^\text{Belle~II} = (387.1 \pm 5.6) \times 10^6$ and $\mathcal{B}(K^{*0}\rightarrow K^+ \pi^-) = 2/3$ assuming isospin conservation.
The signal branching fraction $\BR(B^0\to K^{*0}\tau\ell) $ is a common parameter of the simultaneous fit.

The signal PDF is modeled using a Johnson function~\cite{Johnson:1949zj}, defined as
\begin{equation}\label{eq:RooJohnson}
    \mathcal{P}_\text{sig}(M_\tau, \mu, \lambda, \gamma, \delta) = 
    \frac{\delta}{\lambda \sqrt{2\pi}}
    \frac{1}{\sqrt{1+(\frac{M_{\tau}-\mu}{\lambda})^2}}
    \exp\left[-\frac{1}{2}\left(\gamma + \delta \sinh^{-1}\left(\frac{M_\tau-\mu}{\lambda}\right)\right)^2\right]
\end{equation}
where $\mu$ is the mean of the Gaussian component, $\lambda$ its width, $\gamma$ the distortion of the distribution to the left/right, and $\delta$ the strength of the Gaussian-like component. 
The parameters describing the signal shape are fixed to
the values obtained from the fit to the simulated samples.
Background events have a smooth distribution in
the $M_\tau$ fit region and are analytically modeled using a second-order polynomial whose coefficients are left free to vary.
To validate the fit before the unblinding of the signal region, pseudo-experiments generated from the data sidebands $M_\tau \in$[1.30,\,1.65]$\cup$[1.90,\,2.30]\gevcc are used. Signal is injected with various branching fraction values and the dataset is fitted with the PDFs described above. No biases are found in these studies.

\section{Systematic uncertainties}
\label{sec:syst}

There are systematic uncertainties in the branching fraction measurements due to the determination of the signal efficiencies, the PDF modeling and the external inputs.

The efficiency of the requirement on $\mathcal{P}_\mathrm{FEI}$ is calibrated using the control channel $B^0 \rightarrow D^- \pi^+$. After  reconstructing the \Btag and the charged pion, we search for the $D$ meson resonance in the recoil mass of the $\Btag\pi$ system. 
Calibration factors, defined as weights to account for data/MC differences, are obtained by comparing the yields in data and simulation, for each \Btag decay mode. In Belle II, the inclusive semileptonic decay $B\rightarrow X \ell \nu$  is also used, and the calibration factors are combined with those from the hadronic control channel~\cite{Belle-II:2020fst}.
The signal efficiencies are corrected using those calibration factors and their associated uncertainties are
taken into account as systematic uncertainties.

We take into account the systematic uncertainty associated with the  corrections to the simulated lepton-identification efficiencies, derived from auxiliary measurements in data using $\jpsi\to \mup\mun$, $\epem\to \ell^+\ell^-\gamma$, and $\epem\to\epem\mup\mun$ events. 
These corrections are obtained as functions of track momentum, polar angle and charge, and  applied to events reconstructed from simulation.
The systematic uncertainty is obtained by varying the correction within their uncertainties and estimating the impact of these variations on the selection efficiency.
A similar method is employed for systematic uncertainty due to hadron identification, using the $D^{*+}\to D^0(\to K^- \pi^+)\pi^+$ decays. 

\begin{figure}[h!]
    \centering
    \includegraphics[width=0.4\linewidth]{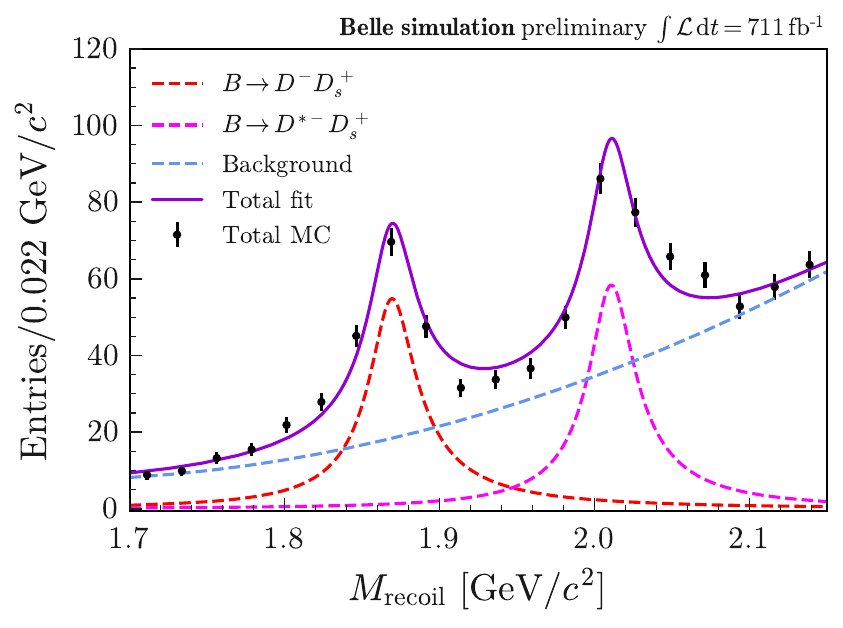}
    \includegraphics[width=0.4\linewidth]{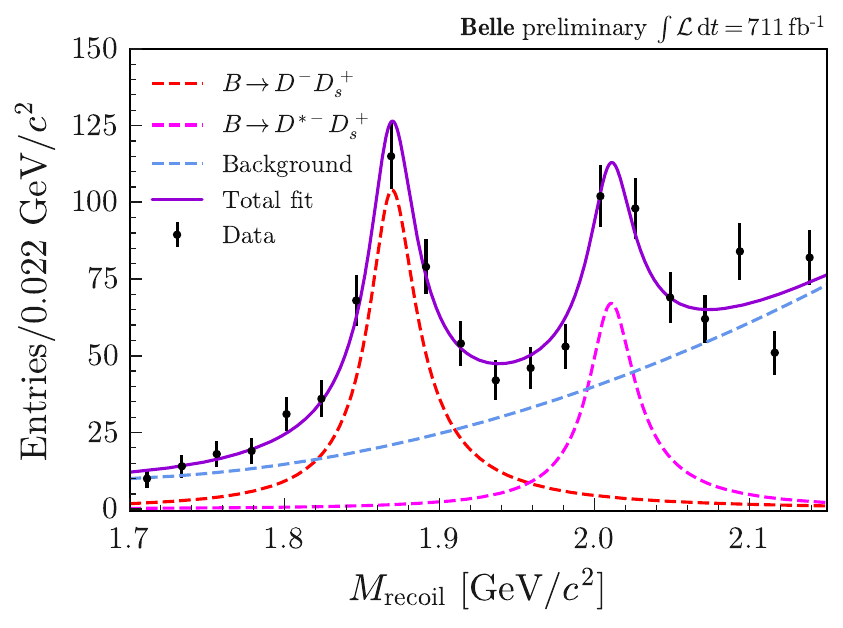}\\
    \includegraphics[width=0.4\linewidth]{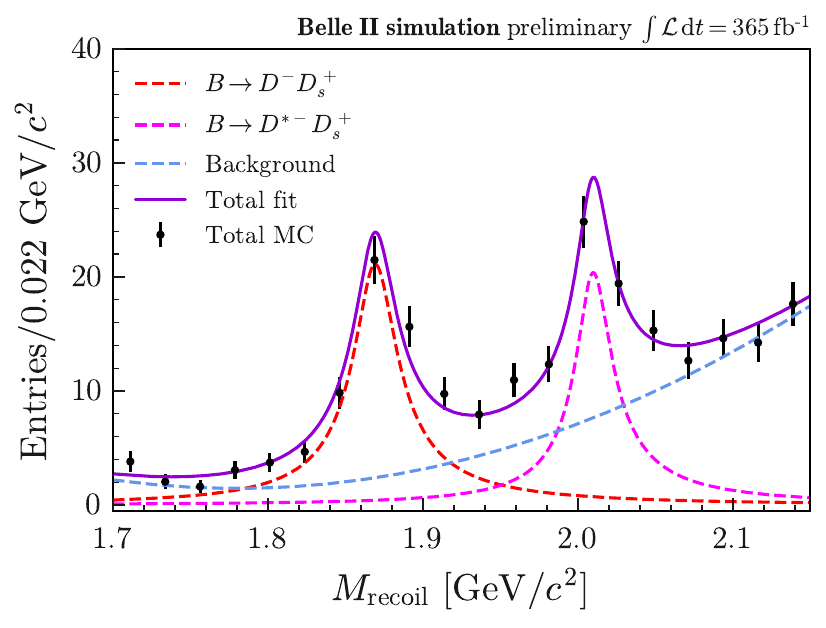}
    \includegraphics[width=0.4\linewidth]{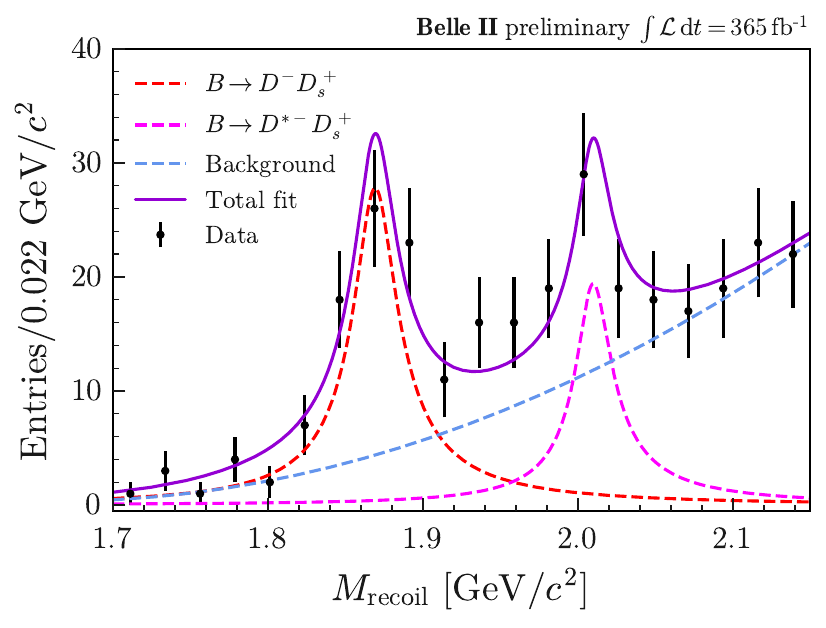}\\
    \caption{Fits to the recoil mass of the \Btag and $D_s^+ $ system in the control channel $B^0\rightarrow D^- D_s^+$ for simulation (left) and data (right). The top plots correspond to Belle data, while the bottom ones are for Belle II data. }
    \label{fig:CC}
\end{figure}

The efficiency of the requirements on the BDT outputs is evaluated using the  $B^0\rightarrow D^- D_s^+(\rightarrow\Kstarz K^+ / \phi \pi^+)$ control sample, with $K^{*0}\rightarrow K^+\pi^-$ and $\phi\rightarrow K^+K^-$. Here, the \Btag is reconstructed in a hadronic channel using the FEI algorithm and the $D_s^+$ is used as a proxy for the $K^{*0}\ell$ system. The $t_\tau$ track is obtained by selecting a random track from the $D^-$ with the correct charge, while other 
$D^-$ decay products are treated as the missing energy.
The selection criteria applied to the control channel are identical to the signal channels when relevant (\Btag, hadron identification, $K^{*0}$ mass window, ROE selection).
In addition, the $K^{*0}K/\phi\pi$ system is selected to be within 20 MeV/$c^2$ of the $D_s$ mass, and in case of $D_s\rightarrow\phi\pi$, the $\phi(\rightarrow KK)$ invariant mass should be within 20 MeV/$c^2$ around the $\phi$ mass.
The recoil mass of the $D^-$ meson is evaluated in the same way as $M_\tau$ for the signal channels.
The $D^-$ recoil mass distributions are shown in Fig.~\ref{fig:CC} for simulation and data for the two datasets.  
A component corresponding to the $D^{*-}$ is also clearly visible.
The uncertainty related to the BDT requirements is obtained by fitting  the  $D^-$ and $D^{*-}$ yields using a Johnson PDF for the signal and  a second-order polynomial for the background, before and after applying the BDT requirement.
The parameters of the Johnson function are fixed to values from fits to the simulation while the background parameters are allowed to float.
Since the control sample has different properties than the signal LFV channels, especially for the ROE, the BDT distributions are expected to differ.
The BDT corresponding to each signal channel is applied to the simulated control channel, and the requirement on the BDT output is set such that its efficiency be the same as the efficiency of the optimized BDT requirement on the signal LFV channel. 
The BDT is also applied on data reconstructed as the control sample and events are selected according to the requirement on the BDT score determined for the control channel. 
The data to MC efficiency ratio $\mathcal{R}^{\text{data/MC}}_{\varepsilon_{\text{BDT}}}$ is measured 
and the assigned uncertainty is symmetrized so that it covers 68.3\% of the area of a Gaussian function with mean $1-\mathcal{R}^{\text{data/MC}}_{\varepsilon_{\text{BDT}}}$ and standard deviation equal to the statistical uncertainty on the ratio.
The corresponding uncertainties on the efficiencies range between 18\% and 34\% and are specified for each mode in Table \ref{tab:syst-summary}. 
Since the data/MC efficiency ratios are compatible with one, no correction is applied to the efficiency and only the systematic uncertainty derived from $\mathcal{R}^{\text{data/MC}}_{\varepsilon_{\text{BDT}}}$ is considered and applied to the signal channels efficiencies.

The difference between data and simulation in the track-reconstruction efficiency in Belle II  is measured in  $e^{+} e^{-} \to \tau^{+} \tau^{-}$ events with $\tau^{-} \to e^{-}\bar{\nu}_{e}\nu_{\tau}$ and $\tau^{-} \to \pi^{-}\pi^{+}\pi^{-}\nu_{\tau}$ to yield a 0.24\% uncertainty per track.
For Belle, a $0.35\%$ per-track uncertainty is assigned using a control sample of
$D^{*+}\to\pi^+ D^0, D^0 \to\pi^+\pi^- K^0_{S}$ decays. 
The uncertainty coming from the limited size of the simulated signal samples is negligible.

The systematic uncertainties related to the signal PDF are obtained from the control sample $B^0\rightarrow D^- D_s^+(\rightarrow\Kstarz K^+ / \phi \pi^+)$, with $K^{*0}\rightarrow K^+\pi^-$ and $\phi\rightarrow K^+K^-$. 
The data are fitted with the nominal PDF, allowing a shift of the mean value $\mu' = \mu + \delta m$ where $\mu$ is fixed to the value fitted in the control channel signal MC and $\delta m$ is free to vary. The largest value between  $\delta m$ and its error is taken as uncertainty on the signal mean, leading to 0.1\% (0.2\%) uncertainty for Belle (Belle II).
For the parameter $\lambda$, corresponding to the width of the signal distribution, we define $\lambda' = \lambda f$ where $\lambda$ is fixed to the value fitted in the control channel signal MC and $f$ is a free scaling factor.
Similarly to the systematic uncertainty of the mean, data is fitted with $\lambda'$, and the maximum of $1-f$ and the error on the fitted value of $f$ is taken as the systematic error on this parameter, leading to 21\% (59\%) uncertainty on the signal resolution for Belle (Belle II). 
The relatively large resulting uncertainty on $\lambda$ is mainly due to the limited size of the control sample.
The tail parameters $\gamma$ and $\delta$ are very sensitive to the background shape. Since no signal is expected, we do not assign any systematic uncertainty on these parameters, and only a dedicated systematic on the background description is estimated.

To assign an uncertainty due to the choice of the background PDF, we fit the data with a third-order polynomial. The variation on the fitted branching fraction with respect to the nominal PDF is taken as a systematic uncertainty.

External inputs to the signal branching fractions determination also introduce systematic uncertainties. Those are related to the number of produced $\Upsilon(4S)$ mesons, $N_{\Upsilon(4S)}^\text{Belle} = (771.6\pm 10.6) \times 10^6$ and $N_{\Upsilon(4S)}^\text{Belle~II} = (387.1 \pm 5.6) \times 10^6$, and the fraction of $\Upsilon(4S)$ decaying into $B^0\bar{B^0}$ pairs, 
 $f^{00}=(48.61\pm0.80)$\% ~\cite{HFLAV:23}. 
The uncertainties on the branching fraction $\mathcal{B}(K^{*0}\to K^+\pi^-)$ and the branching fraction of $\tau$ to one and three prongs are negligible.

A summary of the systematic uncertainties is given in  Table~\ref{tab:syst-summary}.

Systematic uncertainties are taken into account in the upper limit measurement by applying a Gaussian constraint to each of the parameters of the branching fraction. The Gaussian constraint uses the nominal value of the parameter as the mean and for the standard deviation the corresponding systematic uncertainty from Table \ref{tab:syst-summary}.
The systematic uncertainty on the background PDF is estimated directly on the branching fraction, which is the parameter of interest of the fit and thus cannot be Gaussian constrained. 
For that reason, the corresponding systematic uncertainty is applied additively to the final upper limit value.
Pseudo-experiments have been used to check that this method is conservative.
For the fit parameters that have a common uncertainty for Belle and Belle II ($f^{00}$, $\mathcal{B}(K^{*0}\rightarrow K^+ \pi^-)$), a single parameter with the appropriate uncertainty is used in the simultaneous fit. 
The other systematic uncertainty sources are assumed to be uncorrelated between Belle and Belle II.

\begin{table}[h!]
    \centering
     \caption{Summary of the systematic uncertainties.  
    The upper part of the table are systematic uncertainties applied to the efficiency, which are combined in the ``Total efficiency" line. The uncertainties expressed as percentages are multiplicative and applied to the corresponding parameter via a Gaussian constraint in the fit.
    The background PDF uncertainties are additive and directly applied to the branching fraction and upper limit.}
    \vspace{0.2cm}
     \begin{tabular}{c|cccccccc}
    \hline 
    Source & \multicolumn{4}{c}{Belle} & \multicolumn{4}{c}{Belle II} \\   
    & \OSe & \SSe & \OSmu & \SSmu & \OSe & \SSe & \OSmu & \SSmu \\
    \hline
    FEI efficiency [\%] & 4.9 & 4.9 & 4.9 & 4.9 & 6.2 & 6.1 & 6.1 & 6.2\\
    Lepton ID efficiency [\%] & 2.0 & 2.4 & 2.2 & 2.2 & 0.7 & 1.1 & 0.7 & 0.6\\
    Hadron ID efficiency [\%] & 1.9 & 2.0 & 1.9 & 2.0 & 3.7 & 3.7 & 3.6 & 3.7\\
    BDT efficiency [\%] & 27 & 21 & 18 & 23 & 29 & 31 & 34 & 31 \\
    Tracking efficiency [\%] & \multicolumn{4}{c}{1.4} & \multicolumn{4}{c}{1.1} \\
     \cdashline{1-9}
    Total efficiency [\%] & 27.6 & 21.8 & 18.9 & 23.7 & 29.8 & 31.8 & 34.7 & 31.7 \\
    Signal PDF $\mu$ [\%] & \multicolumn{4}{c}{0.1} & \multicolumn{4}{c}{0.2}  \\
    Signal PDF $\lambda$ [\%] & \multicolumn{4}{c}{21} & \multicolumn{4}{c}{59} \\
    $N_{\Upsilon(4S)}$ [\%] & \multicolumn{4}{c}{1.4} & \multicolumn{4}{c}{1.6} \\
    $f^{00}$ [\%] & \multicolumn{8}{c}{0.8} \\ 
    \hline
    Background PDF ($\times10^{-5}$) & 0.11 & 0.28 & 0.09 & 0.02 & 0.11 & 0.28 & 0.09 & 0.02 \\
    \hline \hline
    Total impact on UL ($\times10^{-5}$) & 0.3 & 0.9 & 0.4 & 0.5 & 0.3 & 0.9 & 0.4 & 0.5 \\
    \hline
    \end{tabular}
    \label{tab:syst-summary}
\end{table}

\section{Results and conclusion}
\label{sec:results}
\begin{figure}[h!]
    \centering
    \includegraphics[width=0.42\linewidth]{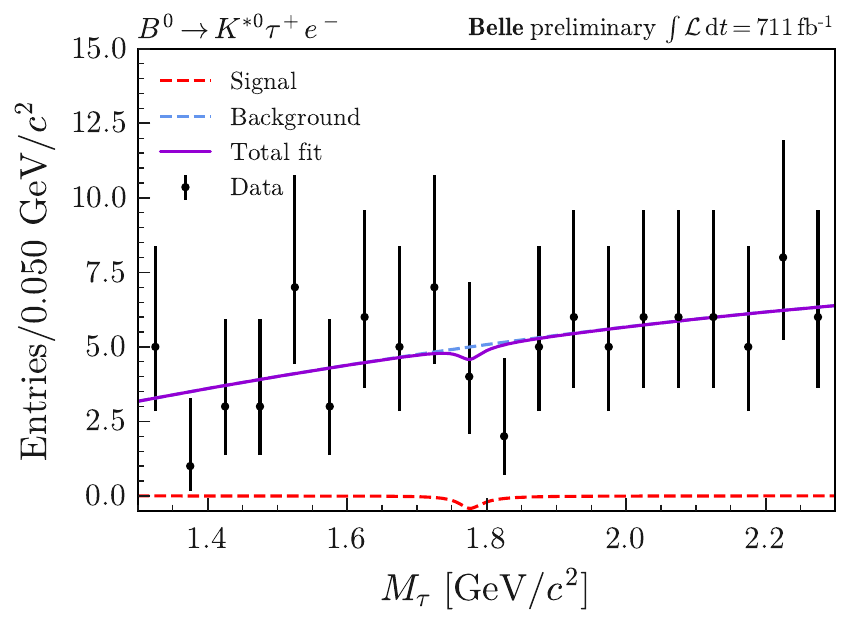}
    \includegraphics[width=0.42\linewidth]{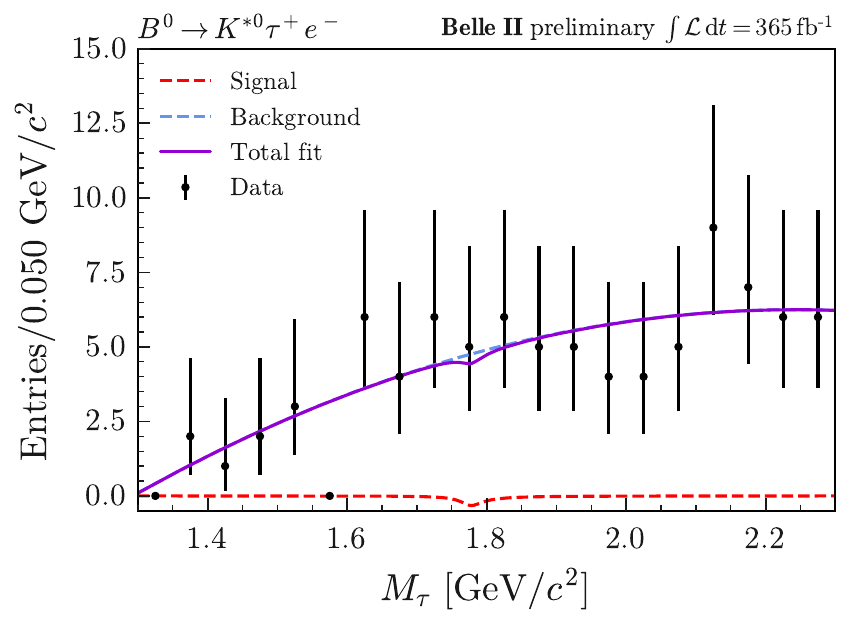}\\
    \includegraphics[width=0.42\linewidth]{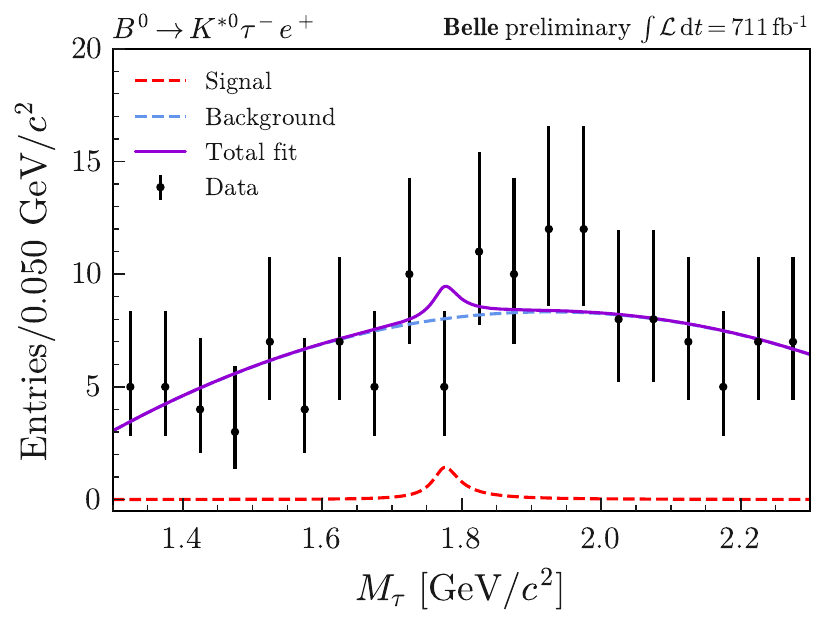}
    \includegraphics[width=0.42\linewidth]{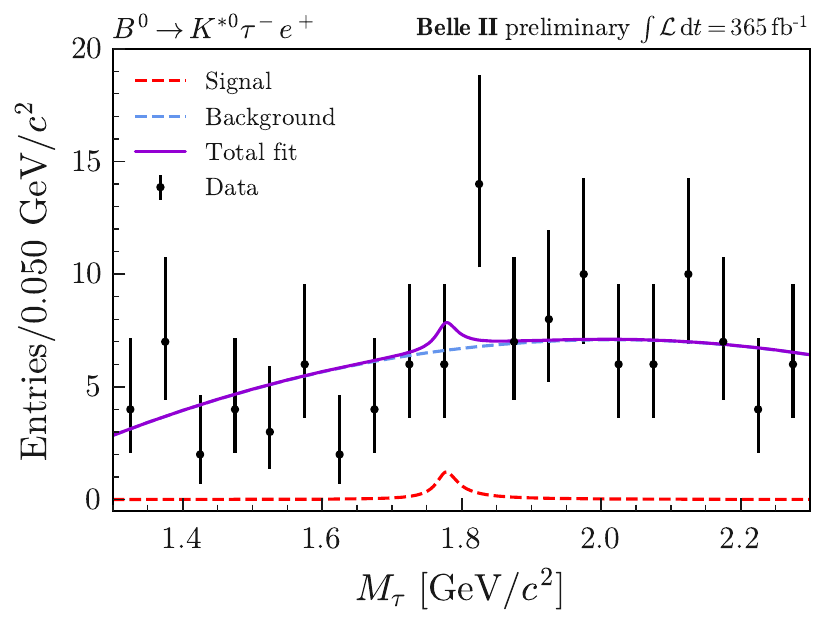}\\
    \includegraphics[width=0.42\linewidth]{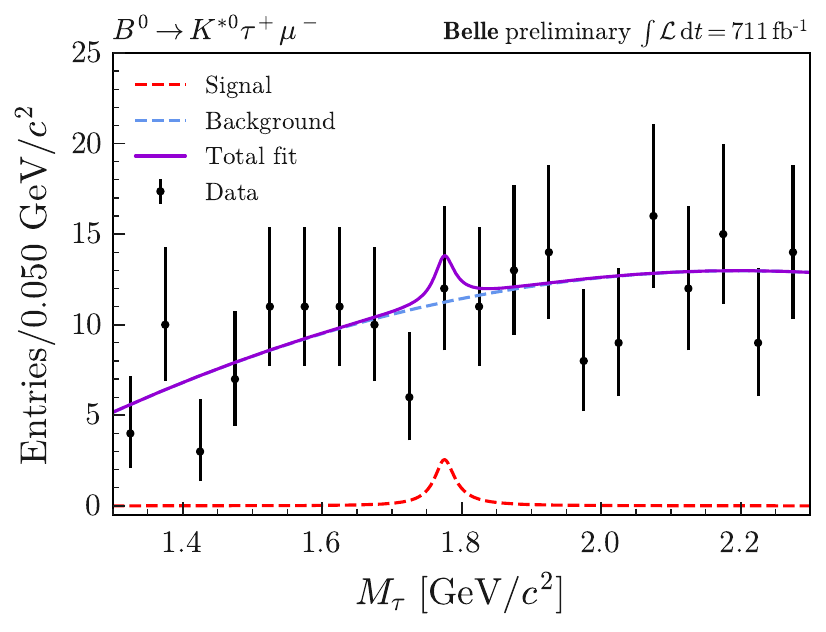}
    \includegraphics[width=0.42\linewidth]{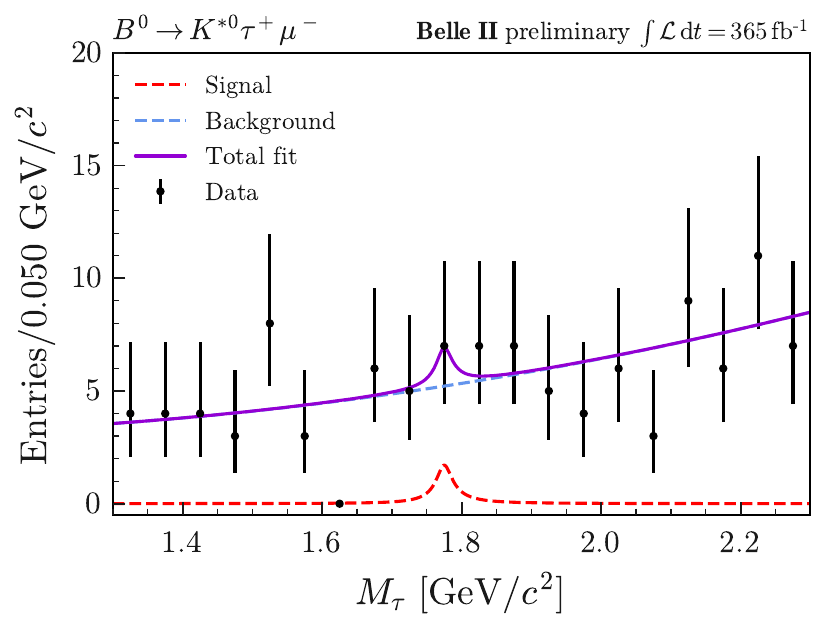}\\
    \includegraphics[width=0.42\linewidth]{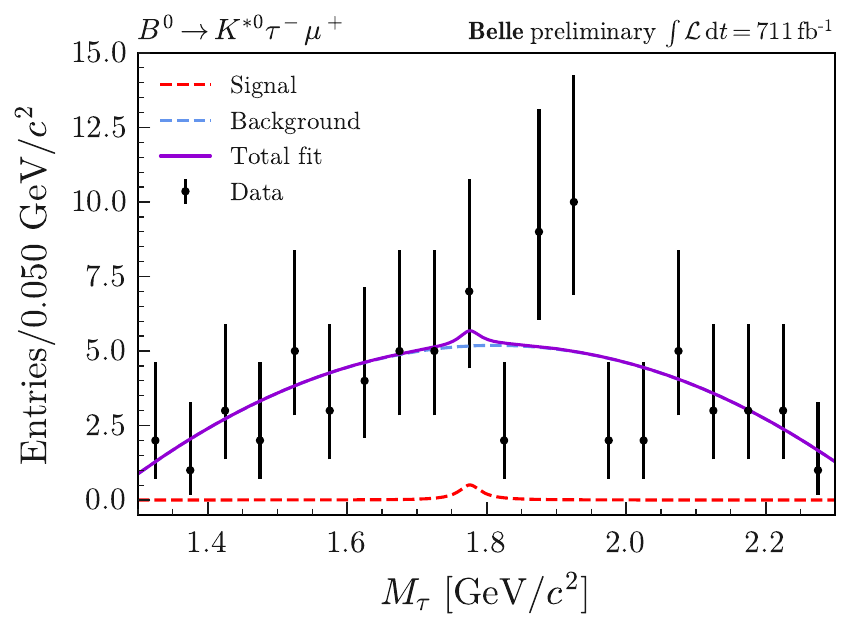}
    \includegraphics[width=0.42\linewidth]{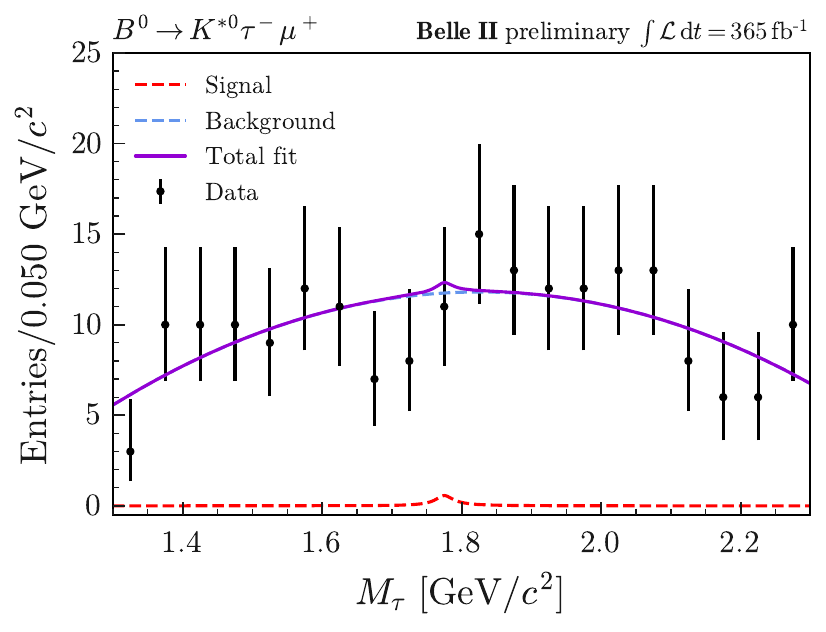}
    \caption{The $M_\tau$ distribution and results of the simultaneous fits for Belle (left) and Belle II (right). The black dots with error bars show the data, the red dash-dotted curve shows the signal component, the blue dashed curve shows the background component, and the purple solid curve shows the global fit. From top to bottom: \OSe, \SSe, \OSmu, \SSmu.}
    \label{fig:totfit}
\end{figure}

The fit results are shown in Fig.~\ref{fig:totfit}
with the corresponding branching fractions ($\mathcal B^\text{fit}$) given in Table~\ref{tab:UL_fin}, where the uncertainties on the fitted branching fractions contain both the statistical and the systematic components.
The fitted number of signal and background in Belle and Belle II are shown in Table~\ref{tab:fit_pars}.
The values for the number of signal are extracted according to expression \eqref{eq:BF} from the efficiencies displayed in Table~\ref{tab:eff} and the fitted branching fractions shown in Table~\ref{tab:UL_fin}.
All fits are validated with pseudo-experiments and no bias is observed.
Since no signal is observed, we set upper limits on the branching fraction using the CLs asymptotic method \cite{Read:2002hq, stat:CLsAsymp}. The observed upper limits ($\mathcal{B}^\text{UL}_\text{obs}$) at 90\% C.L.\ are given in Table~\ref{tab:UL_fin} together with the expected ones ($\mathcal{B}^\text{UL}_\text{exp}$).

The expected limits are computed from a fit to the data sidebands, assuming a number of observed events in the signal region equal to that extrapolated from the sidebands.
\\
The upper limits on the branching fraction at 90\% C.L.\ are:
\begin{align}\label{eq:results}
    \mathcal{B}(B^0\rightarrow K^{*0}\tau^+e^-) < 2.9 \times 10^{-5}, \nonumber\\
    \mathcal{B}(B^0\rightarrow K^{*0}\tau^-e^+) < 6.4 \times 10^{-5}, \nonumber\\
    \mathcal{B}(B^0\rightarrow K^{*0}\tau^+\mu^-) < 4.2 \times 10^{-5}, \nonumber\\
    \mathcal{B}(B^0\rightarrow K^{*0}\tau^-\mu^+) < 5.6 \times 10^{-5}.
\end{align}
Those results are the first search for $B^0\rightarrow K^{*0}\tau\ell$ LFV decays at $e^+e^-$ $B$ factories.
They are the most stringent upper limits to date for the electron modes.

Finally, we also provide upper limits considering two specific BSM scenarios instead of the phase space signal model. 
To do so, we reweight the signal efficiency according to a left-handed model where the Wilson coefficients deviate from the SM by $\Delta C^{\tau\ell}_9 = -\Delta C^{\tau\ell}_{10} \neq 0$ and a scalar model where $\Delta C^{\tau\ell}_S \neq 0$ \cite{Becirevic:2024vwy}.  
In both cases, all other deviations from the SM are 0.
The limits  evaluated with those new efficiencies  are shown in Table \ref{tab:UL_fin}. 
They are similar or higher than the phase space model by up to 19\%.

The results presented in this paper are available on HEPData \cite{HEPData:KstTauEll}.

\begin{table}[!hbp]
    \centering
    \caption{Fitted values of the number of signal ($N_{\text{sig}}$) and background ($N_{\text{bkg}}$) in the Belle and Belle II datasets. The value for $N_{\text{sig}}$ is extracted from the efficiency and fitted branching fraction.}
    \vspace{0.2cm}
\begin{tabular}{cc|cccc}     
\hline
&& $OSe$ & $SSe$ & $OS\mu$ & $SS\mu$\\
\hline
\multirow{2}{*}{Belle} & $N_{\text{sig}}$ & $-0.6\pm3.3$ & $2.2\pm5.2$ & $2.8\pm4.7$ & $0.6\pm3.2$  \\
& $N_{\text{bkg}}$ & $99.5\pm10.4$ & $140.0\pm12.7$ & $207.4\pm15.0$ & $76.4\pm9.3$   \\
\multirow{2}{*}{Belle II} & $N_{\text{sig}}$ &  $-0.5\pm2.8$  & $1.7\pm3.9$ & $1.6\pm2.7$ & $0.6\pm3.3$ \\
& $N_{\text{bkg}}$ & $86.5\pm9.8$  & $120.1\pm11.8$ & $109.1\pm10.9$ & $198.4\pm14.5$ \\
\hline
\end{tabular}
\label{tab:fit_pars}
\end{table}


\begin{table}[h!]
    \centering
     \caption{Measured branching fractions and observed (expected) upper limits at 90\% CL on the four $B^0\to K^{*0}\tau^\pm \ell^\pm$ decays.}
     \vspace{0.2cm}
    \resizebox{\textwidth}{!}{\begin{tabular}{c|cc|cc}
    \hline
    Decay & $\mathcal{B}^\text{fit}$ ($\times10^{-5}$) & $\mathcal{B}^\text{UL}_\text{obs(exp)}$ ($\times10^{-5}$) & $\mathcal{B}^\text{UL}_\text{left}$ ($\times10^{-5}$) & $\mathcal{B}^\text{UL}_\text{scalar}$ ($\times10^{-5}$)  \\
     \hline
    \OSe: $B^0\rightarrow K^{*0}\tau^+e^-$ & $-0.24\pm1.46$ & 2.9 (2.8) & 3.0 & 3.2\\
    \SSe: $B^0\rightarrow K^{*0}\tau^-e^+$ & $1.17\pm2.77$ & 6.4 (4.4) & 7.3 & 7.6 \\
    \OSmu: $B^0\rightarrow K^{*0}\tau^+\mu^-$ & $1.07\pm1.80$ & 4.2 (3.0) & 4.1 & 4.3 \\
    \SSmu: $B^0\rightarrow K^{*0}\tau^-\mu^+$ & $0.48\pm2.61$ & 5.6 (5.5) & 6.0 & 6.4 \\
    \hline
    \end{tabular}}
    \label{tab:UL_fin}
\end{table}

\FloatBarrier

This work, based on data collected using the Belle II detector, which was built and commissioned prior to March 2019,
and data collected using the Belle detector, which was operated until June 2010,
was supported by
Higher Education and Science Committee of the Republic of Armenia Grant No.~23LCG-1C011;
Australian Research Council and Research Grants
No.~DP200101792, 
No.~DP210101900, 
No.~DP210102831, 
No.~DE220100462, 
No.~LE210100098, 
and
No.~LE230100085; 
Austrian Federal Ministry of Education, Science and Research,
Austrian Science Fund (FWF) Grants
DOI:~10.55776/P34529,
DOI:~10.55776/J4731,
DOI:~10.55776/J4625,
DOI:~10.55776/M3153,
and
DOI:~10.55776/PAT1836324,
and
Horizon 2020 ERC Starting Grant No.~947006 ``InterLeptons'';
Natural Sciences and Engineering Research Council of Canada, Compute Canada and CANARIE;
National Key R\&D Program of China under Contract No.~2022YFA1601903,
National Natural Science Foundation of China and Research Grants
No.~11575017,
No.~11761141009,
No.~11705209,
No.~11975076,
No.~12135005,
No.~12150004,
No.~12161141008,
No.~12475093,
and
No.~12175041,
and Shandong Provincial Natural Science Foundation Project~ZR2022JQ02;
the Czech Science Foundation Grant No.~22-18469S 
and
Charles University Grant Agency project No.~246122;
European Research Council, Seventh Framework PIEF-GA-2013-622527,
Horizon 2020 ERC-Advanced Grants No.~267104 and No.~884719,
Horizon 2020 ERC-Consolidator Grant No.~819127,
Horizon 2020 Marie Sklodowska-Curie Grant Agreement No.~700525 ``NIOBE''
and
No.~101026516,
and
Horizon 2020 Marie Sklodowska-Curie RISE project JENNIFER2 Grant Agreement No.~822070 (European grants);
L'Institut National de Physique Nucl\'{e}aire et de Physique des Particules (IN2P3) du CNRS
and
L'Agence Nationale de la Recherche (ANR) under Grant No.~ANR-21-CE31-0009 (France);
BMBF, DFG, HGF, MPG, and AvH Foundation (Germany);
Department of Atomic Energy under Project Identification No.~RTI 4002,
Department of Science and Technology,
and
UPES SEED funding programs
No.~UPES/R\&D-SEED-INFRA/17052023/01 and
No.~UPES/R\&D-SOE/20062022/06 (India);
Israel Science Foundation Grant No.~2476/17,
U.S.-Israel Binational Science Foundation Grant No.~2016113, and
Israel Ministry of Science Grant No.~3-16543;
Istituto Nazionale di Fisica Nucleare and the Research Grants BELLE2,
and
the ICSC – Centro Nazionale di Ricerca in High Performance Computing, Big Data and Quantum Computing, funded by European Union – NextGenerationEU;
Japan Society for the Promotion of Science, Grant-in-Aid for Scientific Research Grants
No.~16H03968,
No.~16H03993,
No.~16H06492,
No.~16K05323,
No.~17H01133,
No.~17H05405,
No.~18K03621,
No.~18H03710,
No.~18H05226,
No.~19H00682, 
No.~20H05850,
No.~20H05858,
No.~22H00144,
No.~22K14056,
No.~22K21347,
No.~23H05433,
No.~26220706,
and
No.~26400255,
and
the Ministry of Education, Culture, Sports, Science, and Technology (MEXT) of Japan;  
National Research Foundation (NRF) of Korea Grants
No.~2016R1-D1A1B-02012900,
No.~2018R1-A6A1A-06024970,
No.~2021R1-A6A1A-03043957,
No.~2021R1-F1A-1060423,
No.~2021R1-F1A-1064008,
No.~2022R1-A2C-1003993,
No.~2022R1-A2C-1092335,
No.~RS-2023-00208693,
No.~RS-2024-00354342
and
No.~RS-2022-00197659,
Radiation Science Research Institute,
Foreign Large-Size Research Facility Application Supporting project,
the Global Science Experimental Data Hub Center, the Korea Institute of
Science and Technology Information (K24L2M1C4)
and
KREONET/GLORIAD;
Universiti Malaya RU grant, Akademi Sains Malaysia, and Ministry of Education Malaysia;
Frontiers of Science Program Contracts
No.~FOINS-296,
No.~CB-221329,
No.~CB-236394,
No.~CB-254409,
and
No.~CB-180023, and SEP-CINVESTAV Research Grant No.~237 (Mexico);
the Polish Ministry of Science and Higher Education and the National Science Center;
the Ministry of Science and Higher Education of the Russian Federation
and
the HSE University Basic Research Program, Moscow;
University of Tabuk Research Grants
No.~S-0256-1438 and No.~S-0280-1439 (Saudi Arabia), and
Researchers Supporting Project number (RSPD2025R873), King Saud University, Riyadh,
Saudi Arabia;
Slovenian Research Agency and Research Grants
No.~J1-9124
and
No.~P1-0135;
Ikerbasque, Basque Foundation for Science,
the State Agency for Research of the Spanish Ministry of Science and Innovation through Grant No. PID2022-136510NB-C33,
Agencia Estatal de Investigacion, Spain
Grant No.~RYC2020-029875-I
and
Generalitat Valenciana, Spain
Grant No.~CIDEGENT/2018/020;
the Swiss National Science Foundation;
The Knut and Alice Wallenberg Foundation (Sweden), Contracts No.~2021.0174 and No.~2021.0299;
National Science and Technology Council,
and
Ministry of Education (Taiwan);
Thailand Center of Excellence in Physics;
TUBITAK ULAKBIM (Turkey);
National Research Foundation of Ukraine, Project No.~2020.02/0257,
and
Ministry of Education and Science of Ukraine;
the U.S. National Science Foundation and Research Grants
No.~PHY-1913789 
and
No.~PHY-2111604, 
and the U.S. Department of Energy and Research Awards
No.~DE-AC06-76RLO1830, 
No.~DE-SC0007983, 
No.~DE-SC0009824, 
No.~DE-SC0009973, 
No.~DE-SC0010007, 
No.~DE-SC0010073, 
No.~DE-SC0010118, 
No.~DE-SC0010504, 
No.~DE-SC0011784, 
No.~DE-SC0012704, 
No.~DE-SC0019230, 
No.~DE-SC0021274, 
No.~DE-SC0021616, 
No.~DE-SC0022350, 
No.~DE-SC0023470; 
and
the Vietnam Academy of Science and Technology (VAST) under Grants
No.~NVCC.05.12/22-23
and
No.~DL0000.02/24-25.

These acknowledgements are not to be interpreted as an endorsement of any statement made
by any of our institutes, funding agencies, governments, or their representatives.

We thank the SuperKEKB team for delivering high-luminosity collisions;
the KEK cryogenics group for the efficient operation of the detector solenoid magnet and IBBelle on site;
the KEK Computer Research Center for on-site computing support; the NII for SINET6 network support;
and the raw-data centers hosted by BNL, DESY, GridKa, IN2P3, INFN, 
PNNL/EMSL, 
and the University of Victoria.

\bibliographystyle{JHEP}
\bibliography{references}

\section*{Additional Material}

\begin{figure}[h!]
    \centering
    \includegraphics[width=0.38\linewidth]{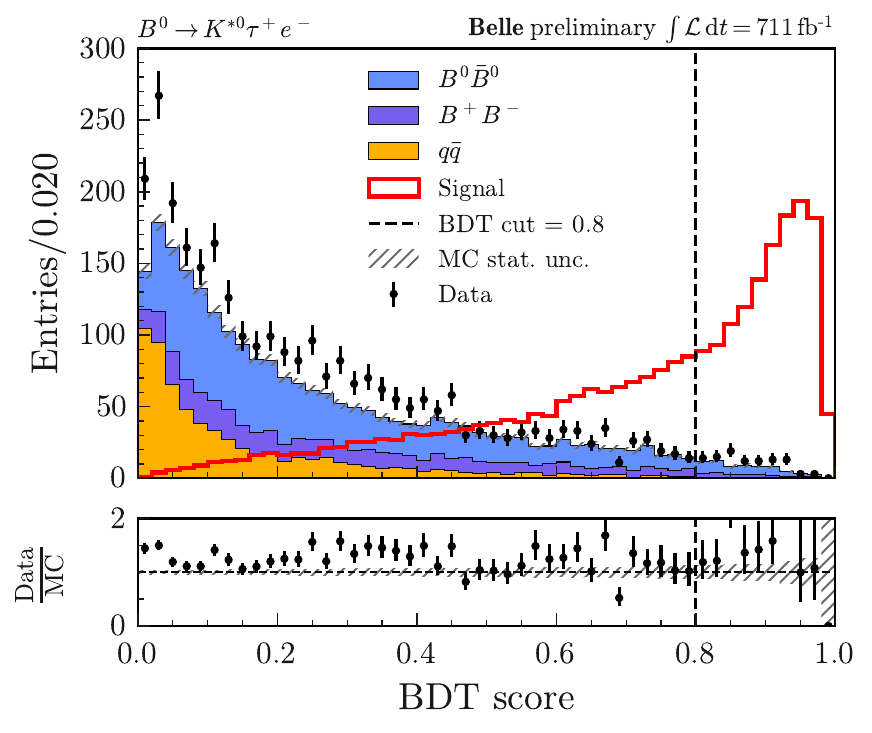}
    \includegraphics[width=0.38\linewidth]{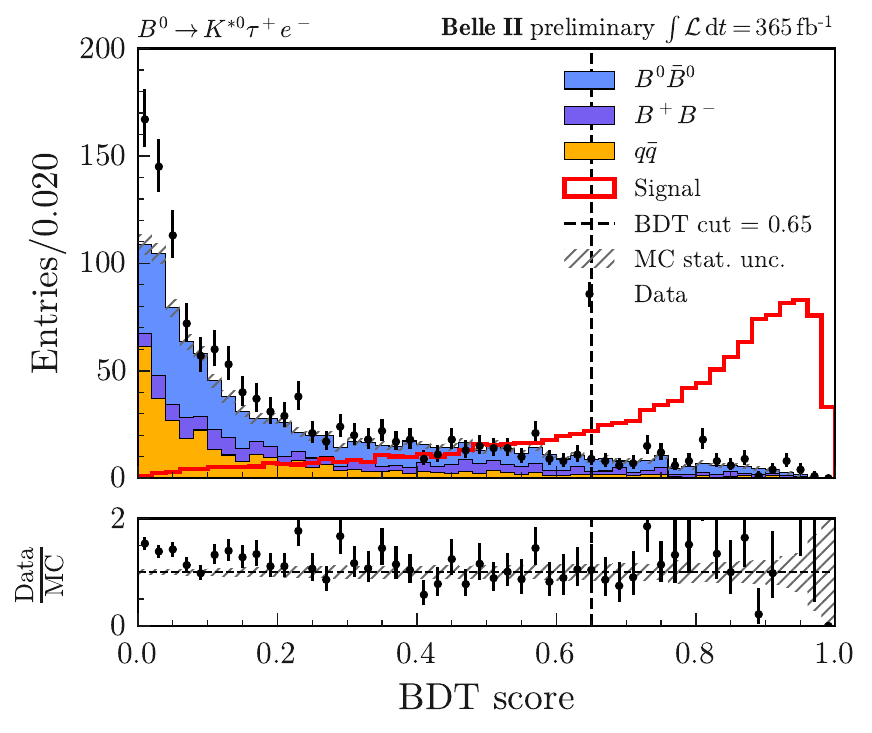}
    \includegraphics[width=0.38\linewidth]{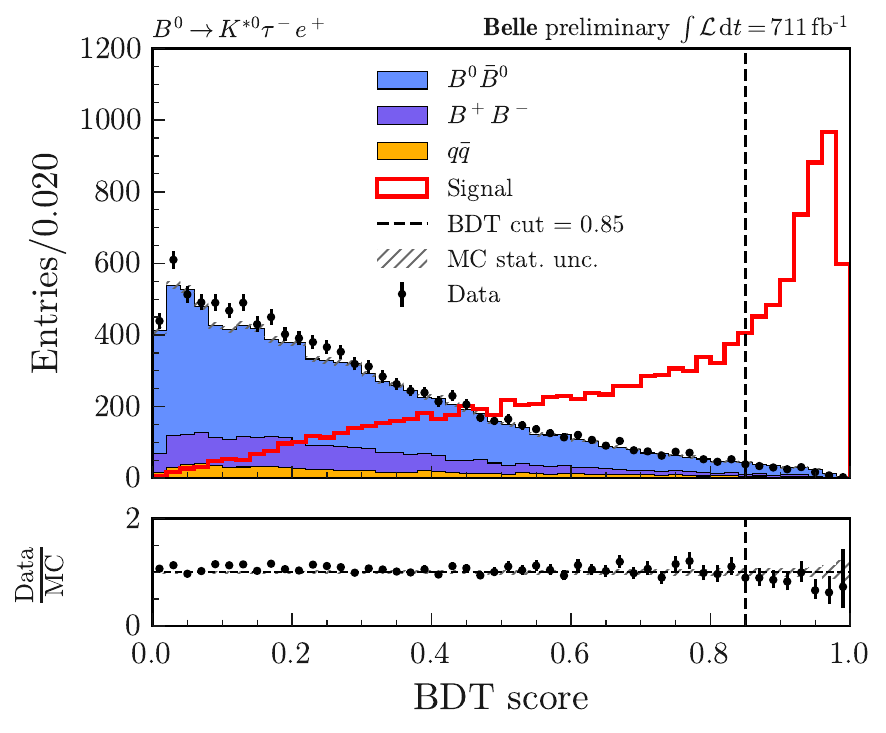}
    \includegraphics[width=0.38\linewidth]{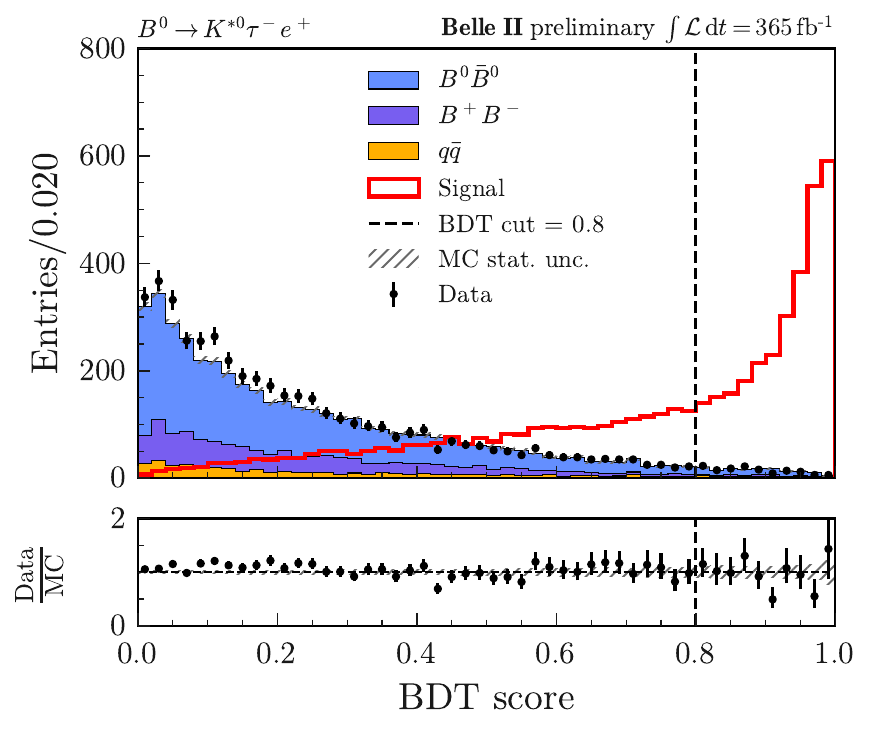}
    \includegraphics[width=0.38\linewidth]{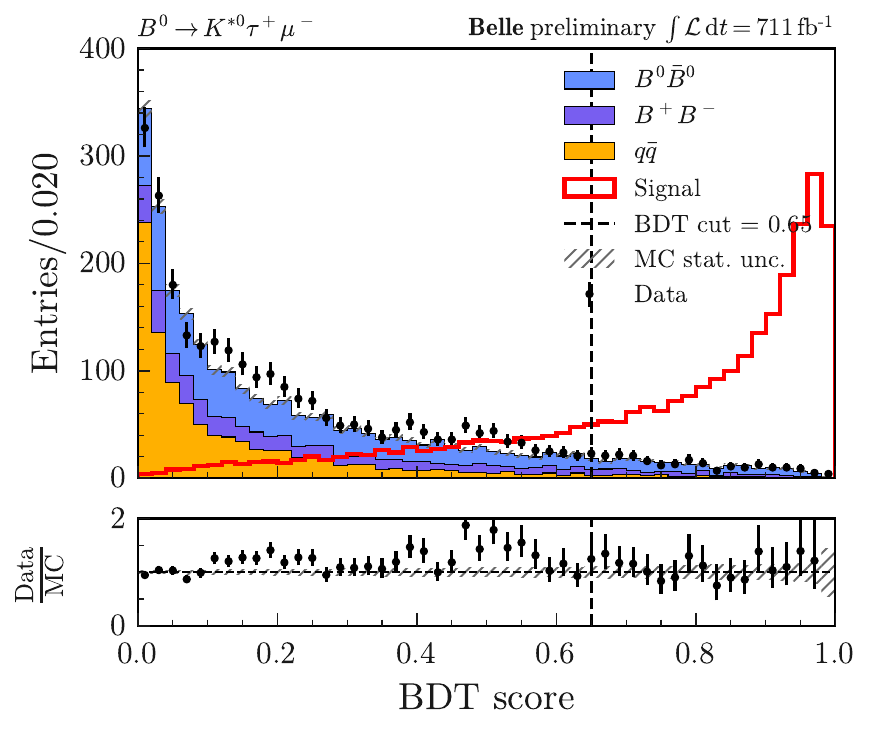}
    \includegraphics[width=0.38\linewidth]{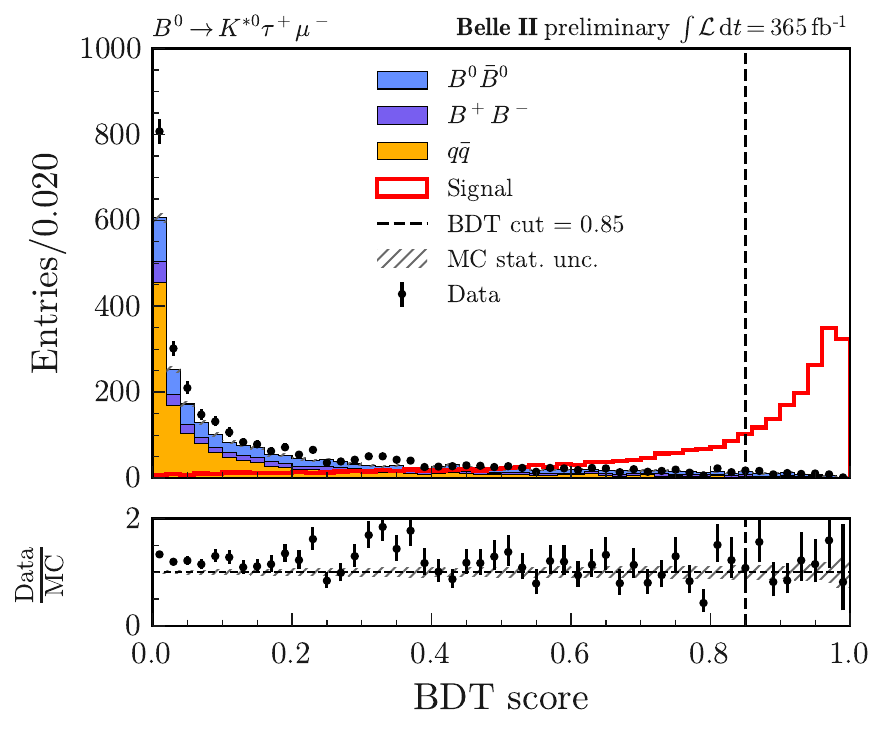}
    \includegraphics[width=0.38\linewidth]{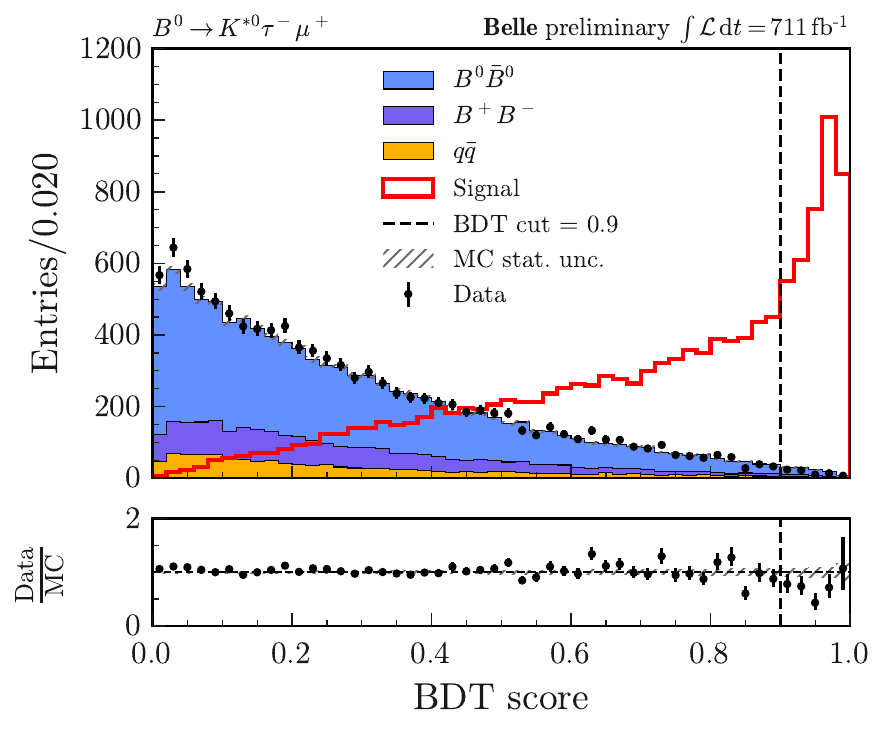}
    \includegraphics[width=0.38\linewidth]{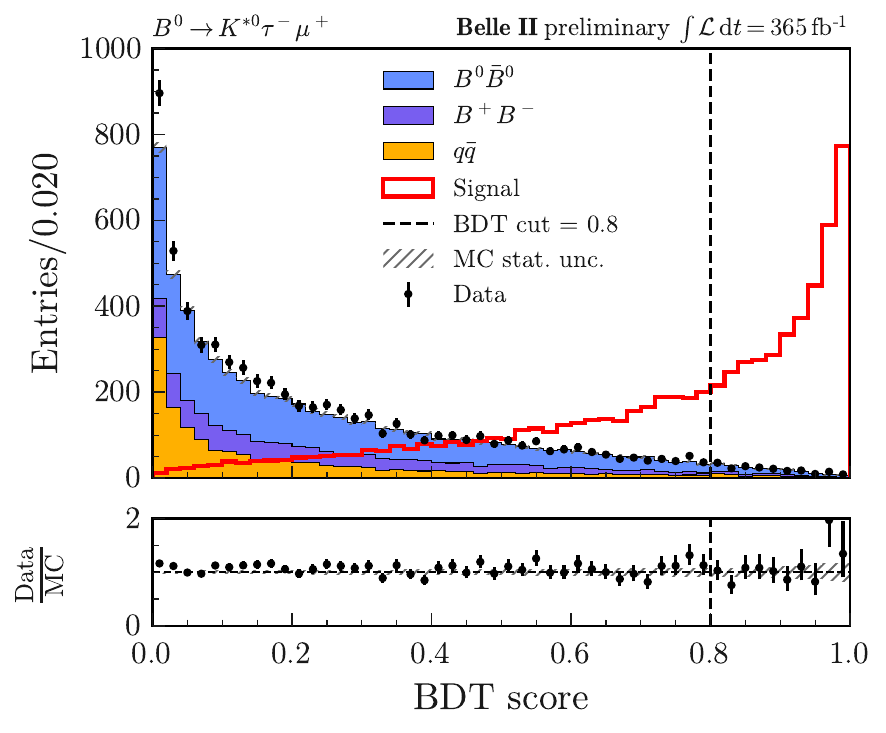}
    \caption{BDT distributions in generic MC simulations (stacked histograms), data sidebands $M_\tau \in [1.0, 1.65[\cup]1.9, 2.5]$\gevcc (black points) and signal MC (red histogram) for Belle (left) and Belle II (right) datasets. The generic MC are corrected for known data/MC mismodelling and scaled to the data luminosity, and signal MC is scaled to the same area as data. The dashed line indicates the value of the BDT cut applied. From top to bottom: \OSe, \SSe, \OSmu, \SSmu.}
    \label{fig:BDT}
\end{figure}

The BDT score distributions of MC simulations, signal MC and sidebands data ($M_\tau \in [1.0, 1.65[\cup]1.9, 2.5]$\gevcc) of the four modes for Belle and Belle II are shown in Figure \ref{fig:BDT}.

\end{document}